\documentclass[twocolumn, prl, superscriptaddress,notitlepage]{revtex4-2}
\usepackage{graphicx}
\usepackage{dcolumn}
\usepackage{amsthm}  
\usepackage{amsmath} 
\usepackage{comment}
\usepackage{bm}
\usepackage[usenames,dvipsnames]{color}
\usepackage[most]{tcolorbox}
\usepackage{multirow}
\usepackage{gensymb}
\usepackage{amssymb}
\usepackage{amsmath}
\usepackage{xcolor}
\usepackage{algorithm}
\usepackage{algorithmic}
\usepackage[normalem]{ulem}
\usepackage{CJK}
\usepackage{comment}
\usepackage{amsfonts}
\usepackage[colorlinks, linkcolor=blue,anchorcolor=blue,citecolor=blue,urlcolor=blue]{hyperref}
\usepackage{amssymb}
\usepackage{pifont}
\usepackage{physics}
\usepackage{natbib}
\usepackage{xcolor}

\usepackage{color,soul}

\begin{document}

\title{Optimal Hamiltonian Parameter Estimation in the Presence of Nuisance Parameters}

\author{Zhiyao Hu }\email[]{zhiyaohu.phys@gmail.com}

\affiliation{Pritzker School of Molecular Engineering, The University of Chicago, Chicago, Illinois, USA}

\author{Haidong Yuan}\email[]{hdyuan@mae.cuhk.edu.hk}
\affiliation{
   Department of Mechanical and Automation Engineering, The Chinese University of Hong Kong, Shatin, Hong Kong}

\author{Liang Jiang}\email[]{liang.jiang@uchicago.edu}
\affiliation{Pritzker School of Molecular Engineering, The University of Chicago, Chicago, Illinois, USA}

\author{Zain H. Saleem}\email[]{zsaleem@anl.gov}
\affiliation{Mathematics and Computer Science Division, Argonne National Laboratory, Lemont, IL, USA}

\begin{abstract}

In many sensing applications, the quantity of interest is not the only unknown—there are also additional unknown parameters, known as nuisance parameters, that affect the precision of estimation. While the ultimate local precision limit for a target parameter is well understood in the absence of nuisance parameters, the problem becomes significantly more challenging when they are present. In this work, we develop a framework for optimal Hamiltonian parameter estimation in the presence of nuisance parameters. We introduce an effective generator that captures the influence of nuisance parameters on the target precision, providing an explicit characterization of the ultimate precision limit for estimating the target parameter. Finally, we provide explicit optimal protocols—including probe state, control, and measurement—that saturate this fundamental limit.
\end{abstract}
\maketitle

\paragraph{Introduction}---
Quantum metrology exploits intrinsically quantum resources, such as coherence and entanglement, to estimate physical quantities with precision beyond classical limits \cite{introhelstrom1969quantum,introholevo2001statistical,introholevo2011probabilistic,qmMaccone_2004_review,qmGiovannetti_2011_review,qmGiovannetti_2006_metrology,qmdegen2017quantum,qmtoth2014quantum,qmpezze2018quantum,polino2020photonic,demille2024quantum,liu2024fully,liu2022optimal,lawrie2019quantum,pirandola2018advances,fujiwara2008fibre,escher2011general,demkowicz2012elusive}. Many applications of quantum metrology can be formulated as Hamiltonian parameter estimation, where an unknown parameter is encoded in the dynamics of a quantum system through a parameter-dependent Hamiltonian \cite{yuan2015optimal,qmpang_2014_metrology,qmpang_optimal_2017,mpyuan_sequential_2016}. Examples include the estimation of magnetic fields \cite{hou2021zero,hou2020minimal,isogawa2025entanglement,maze2008nanoscale,taylor2008high}, frequencies \cite{hou2021super,isogawa2026approaching}, forces \cite{gilmore2021quantum,novikov2025hybrid}, phases \cite{cao2025joint,li2023improving,zaporski2025quantum}, and gradients\cite{zhang2026distributed}. 

For the estimation of a single unknown parameter, the ultimate local precision is governed by the quantum Cramér-Rao bound, with the quantum Fisher information quantifying the maximum information that can be extracted from the probe state. For unitary dynamics, this bound has a particularly simple and operational form: the maximal quantum Fisher information is determined by the spectral width of the generator associated with the parameter \cite{qmGiovannetti_2006_metrology}. This generator-based picture not only characterizes the fundamental precision limit, but also directly guides the construction of optimal probe states, coherent controls, and measurements \cite{yuan2015optimal,yuan2017quantum, qmpang_optimal_2017,qmpang_2014_metrology}. Consequently, single-parameter Hamiltonian sensing is now supported by a well-developed and largely constructive theory.

In realistic sensing scenarios, however, the parameter of interest is rarely the only unknown quantity affecting the measurement outcome. Experimental systems often contain additional uncertain parameters that influence the probe dynamics but are not themselves the primary quantities of interest. These are commonly referred to as nuisance parameters \cite{suzuki2020quantum}. Examples include unknown detunings, imperfectly calibrated coupling constants, background fields, phase offsets, or environmental parameters. The presence of nuisance parameters makes the sensing problem intrinsically multiparameter \cite{matsumoto2002new,kahn2009local,li2016fisher,pezze2017optimal,zhu2018universally,albarelli2019evaluating,yang2019optimal,yang2019attaining,sidhu2021tight,crowley2014tradeoff,vidrighin2014joint,yue2014quantum,zhang2014quantum,ragy2016compatibility,suzuki2016explicit,chen2017maximal,liu2017control,roccia2017entangling,chen2019optimal,razavian2020quantumness,sidhu2020geometric,candeloro2021properties,lu2021incorporating,2005asymptotic,zhu2018universally,vidrighin2014joint,guhne2023colloquium,isogawa2026approaching,isogawa2025entanglement,hu2025optimal}. Even though these parameters are not themselves of interest, they can strongly degrade the achievable precision for the target parameter. In this setting, the relevant precision limit is not determined solely by the quantum Fisher information associated with the target parameter \cite{abbasgholinejad2026multiparameter,qneldredge2018optimal}. Rather, it depends on the full quantum Fisher information matrix and, in particular, on the correlations between the target response and the nuisance responses. The effective precision is determined by the part of the target response that cannot be reproduced by variations of the nuisance parameters. As a result, a protocol that is optimal in an idealized single-parameter model may become suboptimal, or may even fail to identify the target parameter, once nuisance parameters are present.

Despite its practical importance, a fully operational theory of optimal Hamiltonian parameter estimation in the presence of nuisance parameters is still lacking.
Although the optimal measurement can, in principle, be determined for a given output state, the joint design of the optimal probe state and control protocol-both essential for achieving the ultimate precision-remains an open challenge. Consequently, the fundamental precision limit for estimating a target Hamiltonian parameter in the presence of nuisance parameters has remained largely unresolved.
Recent work established an optimal scheme for estimating a function of multiple parameters encoded in a time-independent Hamiltonian \cite{abbasgholinejad2026multiparameter}; however, the corresponding problem for general time-dependent Hamiltonians remains unresolved.

In this work, we develop a general framework for optimal Hamiltonian parameter estimation in the presence of nuisance parameters. We show that the ultimate precision for estimating the target parameter is governed by an effective generator, obtained by removing from the target generator the response attributable to nuisance-parameter variations. This effective generator isolates the component of the target response that cannot be reproduced by changes in the nuisance parameters, thereby extending the familiar single-parameter generator formalism to sensing problems with nuisance parameters. Crucially, the effective generator also enables the explicit construction of optimal protocols—including the probe state, control strategy, and measurement—that collectively attain the ultimate precision limit. Finally, we demonstrate the scope and utility of our framework through several representative applications.

A typical quantum-metrology protocol consists of three stages: preparation of an initial probe state \(\lvert\Psi_0\rangle\), controlled evolution encoding the unknown parameters \(\mathbf{x}=(x_1,\ldots,x_N)^T\), and a final measurement on the output state
\[
\lvert\Psi_{\mathbf{x}}(T)\rangle
=
U_{\mathrm{tot}}(T)\lvert\Psi_0\rangle,
\]
where \(U_{\mathrm{tot}}(T)\) is generated by a general time-dependent Hamiltonian $H(\mathbf x,t)=H_0(\mathbf x,t)+H_C(t)$, here
$H_0(\mathbf x,t)$ is the free Hamiltonian that contains the unknown parameters and $H_C(t)$ is the control Hamiltonian that can be externally tuned. For any locally unbiased estimator, the quantum Cramér-Rao bound (QCRB) gives
\begin{equation}
    \mathrm{Cov}(\hat{\mathbf x})
    \ge
    \frac{1}{\nu}
    J_{\mathbf x}^{-1}(T),
    \label{eq:x_qcrb}
\end{equation}
where $\mathrm{Cov}(\hat{\mathbf x})$ is the covariance matrix, \(\nu\) is the number of independent measurements, $J_{\mathbf x}(T)$ is the quantum Fisher information matrix (QFIM) \cite{qcrbfisher1925theory,qcrbrao1992information,introhelstrom1969quantum,qcrbrao1992information,qmCaves_1994_Bures}.
For a pure probe undergoing unitary evolution, the \(jk\)-th entry of the QFIM can be obtained as
\begin{equation}
    [J_{\mathbf x}(T)]_{jk}
    =
    4\,\mathrm{Re}
    \left[
    \langle S_{x_j}(T)S_{x_k}(T)\rangle
    -
    \langle S_{x_j}(T)\rangle
    \langle S_{x_k}(T)\rangle
    \right],
    \label{eq:Jx_definition}
\end{equation}
where 
\begin{align}
    \label{eq:generator}
    S_{x_j}(T)
   &=
    iU_{\rm tot}^\dagger(T)
    \partial_{x_j}U_{\rm tot}(T) \nonumber \\
    &=\int_0^T U_{\rm tot}^\dagger(t)\partial_x H_0(x,t)U_{\rm tot}(t) dt
\end{align}
is the generator associated with \(x_j\) \cite{qmpang_2014_metrology,qmpang_optimal_2017,hu2024control}. In particular, the diagonal
entries reduce to the variances of the corresponding generator, $J_{x_j}(T)
    =4\Delta^2 S_{x_j}(T).$
Here all expectation values are evaluated
with respect to the initial probe state. 

We now consider the estimation of a target parameter $ \theta_1$ in presence of nuisance parameters $\boldsymbol\theta_{\rm n}
    =
    (\theta_2,\ldots,\theta_N)^T$, which is related to the original parameters by an invertible linear transformation \(\boldsymbol{\theta}
    =
    (\theta_1,\boldsymbol{\theta}_{\rm n}^T)^T
    =
    A\mathbf x\).
Under this reparametrization, the QFIM
transforms as \cite{paris2009quantum}
\begin{equation}
    J_{\boldsymbol\theta}(T)
    =
    \left(
    \frac{\partial \mathbf x}
    {\partial \boldsymbol\theta}
    \right)^T
    J_{\mathbf x}(T)
    \left(
    \frac{\partial \mathbf x}
    {\partial \boldsymbol\theta}
    \right)
    =
    A^{-T}
    J_{\mathbf x}(T)
    A^{-1}.
    \label{eq:Jtheta_transform}
\end{equation}

When $\boldsymbol\theta_{\rm n}$ are known, the precision for estimating \(\theta_1\) is determined by the corresponding diagonal QFI, \(\bigl[J_{\boldsymbol{\theta}}(T)\bigr]_{11}\). When they are unknown, however, correlations between the target and nuisance responses reduce the information that can be attributed uniquely to \(\theta_1\). For a nonsingular QFIM, the relevant precision bound is instead determined by the corresponding diagonal element of the inverse QFIM \cite{liu2020quantum,suzuki2020quantum}:
\begin{equation}
    \mathrm{Var}(\hat\theta_1)
    \ge
    \frac{1}{\nu}
    \left[
    J_{\boldsymbol\theta}^{-1}(T)
    \right]_{11}
    :=
    \frac{1}{\nu J_{\theta_1|\boldsymbol\theta_{\rm n}}(T)},
    \label{eq:theta1_qcrb_inverse}
\end{equation}
where we write the QFIM in blocks 
\begin{equation} J_{\boldsymbol\theta}(T) = \begin{pmatrix} J_{\theta_1}(T) 
& J_{\theta_1\boldsymbol\theta_{\rm n}}(T) 
\\ J_{\boldsymbol\theta_{\rm n}\theta_1}(T) & J_{\boldsymbol\theta_{\rm n}}(T) \end{pmatrix}, \end{equation}
and \begin{equation} J_{\theta_1|\boldsymbol\theta_{\rm n}}(T) = J_{\theta_1}(T) - J_{\theta_1\boldsymbol\theta_{\rm n}}(T) J_{\boldsymbol\theta_{\rm n}}^{-1}(T) J_{\boldsymbol\theta_{\rm n}\theta_1}(T)\label{eq:general_schur}. \end{equation}
The second term in Eq.(\ref{eq:general_schur}) quantifies the loss of target information caused by responses that are indistinguishable from variations in the nuisance parameters. 

Although Eq.~\eqref{eq:theta1_qcrb_inverse} provides a formal precision bound for a fixed output state, it offers limited insight into the ultimate sensitivity achievable through optimization of the full sensing protocol. In particular, it depends on the entire QFIM and does not directly reveal how the nuisance response should be removed at the generator level. Nor does it provide a constructive prescription for choosing the probe state and control operations that maximize the effective QFI. While a locally optimal measurement can, in principle, be determined once the output state is fixed, jointly optimizing the probe, control, and measurement remains a substantially more challenging problem. Consequently, a general characterization of the ultimate precision attainable in Hamiltonian parameter estimation with nuisance parameters—and an explicit protocol that achieves it-has remained elusive.

To address this gap, we develop a framework to characterize the ultimate limit for the Hamiltonian parameter estimation in the presence of nuisance parameters and provide explicit protocols to saturate it. At the core of our approach is an effective generator that isolates the target response from variations induced by the nuisance parameters. This formulation yields a transparent characterization of the ultimate precision and enables the constructive design of optimal probe states, control strategies, and measurements.

We begin with the estimation of a single target parameter, $\theta_1 = w_1x_1+w_2x_2$, in the presence of one nuisance parameter  $\theta_2=w_2x_1-w_1x_2 $, where $w_1^2+w_2^2=1$. The choice of nuisance coordinate is made for convenience and does not
affect the precision of the target parameter (see the Supplemental Material for details).
The generators associated with $\theta_1$ and $\theta_2$ are 
\begin{align}
   S_{\theta_1}(T)&=w_1S_{x_1}(T)+w_2S_{x_2}(T),\nonumber\\
    S_{\theta_2}(T)
    &=
    w_2S_{x_1}(T)-w_1S_{x_2}(T).
    \label{eq:theta1_theta2_generators}   
\end{align}
By using Eq.(\ref{eq:Jx_definition}), we can rewrite Eq.(\ref{eq:general_schur}) as
\begin{align}\label{eq:schur_covariance_form}
  J_{\theta_1|\theta_2}(T)
    &=
    4
    \left[
    \Delta^2 S_{\theta_1}(T)
    -
    \frac{
    \mathrm{Cov}^2(S_{\theta_1}(T),S_{\theta_2}(T))
    }{
    \Delta^2 S_{\theta_2}(T)
    }
    \right]
\end{align}
where the symmetrized covariance is defined as
\begin{equation}
\operatorname{Cov}(A,B)
=
\frac{1}{2}
\langle AB+BA\rangle
-
\langle A\rangle\langle B\rangle.
\end{equation}
As shown in the Supplemental Material, Eq.~\eqref{eq:schur_covariance_form} admits an equivalent variational representation
\begin{equation}\label{eq:schur_generator_form}
J_{\theta_1|\theta_2}(T)=4
    \min_{\beta\in\mathbb R}
    \Delta^2
    \left(
        S_{\theta_1}^{\rm eff}(\beta,T)
    \right),
    \end{equation}
where $ S_{\theta_1}^{\rm eff}(\beta,T)
    =
    S_{\theta_1}(T)-\beta S_{\theta_2}(T)$
is the effective generator for the target parameter. 

For any Hermitian operator 
$A$, its variance is bounded by one quarter of the squared spectral seminorm,
\begin{equation}
\Delta^2A
\le
\frac14|A|_s^2,
\label{eq:variance_spectral_bound}
\end{equation}
where
\(|A|_s
=
\lambda_{\max}(A)-\lambda_{\min}(A)\) with $\lambda_{\max(\min)}(A)$ being the maximal (minimal) eigenvalue of the operator $A$.
Applying Eq.~\eqref{eq:variance_spectral_bound} to the effective generator yields 
\begin{align}
J_{\theta_1\mid\theta_2}(T)
=
4\min_{\beta\in\mathbb{R}}
\Delta^2
\left[
S_{\theta_1}^{\mathrm{eff}}(\beta,T)
\right]
\le
\min_{\beta\in\mathbb{R}}
\left|
S_{\theta_1}^{\mathrm{eff}}(\beta,T)
\right|_s^2.
\label{eq:effective_generator_spectral_bound}
\end{align}
We next derive an upper bound on $\left|
S_{\theta_1}^{\mathrm{eff}}(\beta,T)
\right|_s$ for arbitrary controlled evolution and subsequently construct an explicit protocol that saturates it.

Using Eq.~\eqref{eq:generator}, the effective generator can be written as
\begin{align}
     S_{\theta_1}^{\rm eff}(\beta,T)
    =
    \int_0^T
    U_{\rm tot}^\dagger(t)
    V_{\theta_1}^{\rm eff}(\beta,t)
    U_{\rm tot}(t)\,dt,  
\end{align}
where  \(V_{\theta_1}^{\rm eff}(\beta,t)=\partial_{\theta_1}H_0(\mathbf x,t)
   -    \beta\,\partial_{\theta_2}H_0(\mathbf x,t)\) is the instantaneous effective generator. For compactness, we may also write
$V_{\theta_j}(t)
=
\partial_{\theta_j}H_0(\mathbf{x},t)$,
so that
$V_{\theta_1}^{\mathrm{eff}}(\beta,t)
=
V_{\theta_1}(t)-\beta V_{\theta_2}(t)$.

The spectral seminorm of the integrated generator satisfies
\begin{align}
    \left|
    S_{\theta_1}^{\rm eff}(\beta,T)
    \right|_s
    &=
    \left|
    \int_0^T
    U_{\rm tot}^\dagger(t)
    V_{\theta_1}^{\rm eff}(\beta,t)
    U_{\rm tot}(t)\,dt
    \right|_s
    \nonumber\\
    &\le
    \int_0^T
    \left|
    U_{\rm tot}^\dagger(t)
    V_{\theta_1}^{\rm eff}(\beta,t)
    U_{\rm tot}(t)
    \right|_s dt
    \nonumber\\
    &=
    \int_0^T
    \left|  
    V_{\theta_1}^{\rm eff}(\beta,t)
    \right|_s dt.\label{eq:time_independent_triangle_bound}
\end{align}
Here, the inequality follows from the triangle inequality for the spectral seminorm, while the final equality follows from its invariance under unitary conjugation. Combining Eqs.~\eqref{eq:effective_generator_spectral_bound} and \eqref{eq:time_independent_triangle_bound}, we obtain
\begin{equation}
    J_{\theta_1|\theta_2}(T)\le
\min_{\beta\in\mathbb{R}}
\left|
S_{\theta_1}^{\mathrm{eff}}(\beta,T)
\right|_s^2
    \le
    \min_{\beta\in\mathbb R}
   \left[\int_0^T \left|
    V_{\theta_1}^{\rm eff}(\beta,t)
    \right|_sdt\right]^2 .
    \label{eq:TI_qfi_bound}
\end{equation}
This bound is independent of the initial probe state and of the applied control. It therefore provides a universal upper bound on the effective QFI attainable in the presence of the nuisance parameter.

The minimization over $\beta$ is convex and can be expressed as the following semi-definite programming (SDP) problem:
\begin{align}
    \min_{\beta,\,a(t),\,b(t)}\quad
    &
    \int_0^T
    [a(t)-b(t)]dt ,
    \label{eq:TD_sdp}\\
    \mathrm{s.t.}\quad
    &
    b(t)\mathbb I
    \preceq
    V_{\theta_1}(t)-\beta V_{\theta_2}(t)
    \preceq
    a(t)\mathbb I ,
    \qquad 0\le t\le T .
    \nonumber
\end{align}
At the optimum, $a(t)$ and $b(t)$ bound the largest and smallest eigenvalues, respectively, of the instantaneous effective generator. Hence,
\begin{equation}
a(t)-b(t)
=
\left|
V_{\theta_1}(t)-\beta V_{\theta_2}(t)
\right|_s.
\end{equation}
For continuous time, Eq.~\eqref{eq:TD_sdp} is a semidefinite optimization problem with a continuum of matrix inequalities. In practice, it can be reduced to a finite-dimensional SDP by discretizing the evolution interval $[0,T]$ into \(M\) segments. Let \(t_k\) denote a representative point in the \(k\)-th segment and \(\Delta t_k\) its duration. The discretized problem is then 
\begin{align}
&\min_{\beta,\,\{a_k,b_k\}}\quad  \sum_{k=0}^{M-1} \Delta t_k (a_k-b_k), \nonumber\\ 
\mathrm{s.t.}\quad & b_k\mathbb I \preceq V_{\theta_1}(t_k)-\beta V_{\theta_2}(t_k) \preceq a_k\mathbb I , \qquad k=0,\ldots,M-1 . \label{eq:TD_discrete_sdp} 
\end{align}
At the optimum, \(a_k\) and \(b_k\) coincide with the largest and smallest eigenvalues, respectively, of \(V_{\theta_1}(t_k)-\beta V_{\theta_2}(t_k)\). Thus, the objective approximates the time integral of the spectral seminorm. As the discretization is refined, the optimal solution converges to that of the continuous-time problem under standard regularity assumptions.

Denoting the optimal $\beta$ obtained from the SDP as \(\beta^\star\), we then have
\begin{equation}
    J_{\theta_1|\theta_2}(T)
    \leq
   \left[\int_0^T \left|
    V_{\theta_1}^{\rm eff}(\beta^\star,t)
    \right|_sdt \right]^2.
    \label{eq:time_independent_qfi_bound}
\end{equation}

\begin{figure}[htbp]
\includegraphics[width=0.5\textwidth]{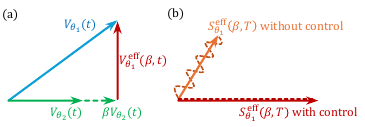}
\caption{
(a) The target and nuisance parameters are encoded through
\(V_{\theta_1}(t)\) and \(V_{\theta_2}(t)\), respectively. Subtracting the nuisance response
\(\beta V_{\theta_2}(t)\) from the target response defines the effective velocity
\(V_{\theta_1}^{\rm eff}(\beta,t)
=V_{\theta_1}(t)-\beta V_{\theta_2}(t)\).
(b) Without control, the instantaneous effective velocities are generally misaligned and partially cancel during the evolution. The optimal control aligns their extremal branches, producing a larger spectral width of the accumulated effective generator and hence a higher effective QFI. Here the orange (red) dashed curve traces the instantaneous effective velocity without (with) control, while the orange (red) solid arrow represents the corresponding accumulated effective generator.}
\label{ci}
\end{figure}

We now show that this bound is attainable. Let \(\lvert v_{\max}(t)\rangle\) and \(\lvert v_{\min}(t)\rangle\) denote eigenvectors of the instantaneous effective generator
\(
V_{\theta_1}^{\mathrm{eff}}(\beta^\star,t)
=
V_{\theta_1}(t)-\beta^\star V_{\theta_2}(t)
\)
associated with its largest and smallest eigenvalues, respectively. An optimal initial probe state is the balanced superposition
\begin{equation}
        \lvert\varphi_0^\star\rangle
    =
    \frac{1}{\sqrt{2}}
    \left(
        \lvert v_{\max}(0)\rangle
        +
        e^{i\phi}
        \lvert v_{\min}(0)\rangle
    \right),
        \label{eq:optimal_initial_state}
\end{equation}
where the relative phase \(\phi\) may be chosen arbitrarily. 
The optimal evolution is to keep the eigenvectors corresponding to the extreme eigenvalues of \(V_{\theta_1}^{\rm eff}(\beta^\star,t)\) aligned for all $t\in[0,T]$. Specifically, the optimal propagator satisfies
\begin{align}\label{eq:optimalcontrol}
    U_{\rm tot}(t)|v_{\max/\min}(0)\rangle
    &=
    |v_{\max/\min}(t)\rangle,
\end{align}
for all $t\in[0,T]$. 
Under this optimal evolution, the extremal eigenvectors of the instantaneous effective generator are mapped onto fixed reference directions in the interaction picture:
    \(U_{\mathrm{tot}}^\dagger(t)
    V_{\theta_1}^{\mathrm{eff}}(\beta^\star,t)
    U_{\mathrm{tot}}(t)
    \lvert v_{\max/\min}(0)\rangle
    =
    \lambda_{\max/\min}(t)
    \lvert v_{\max/\min}(0)\rangle,\nonumber \)
where \(\lambda_{\max/\min}(t)\) are the largest and smallest eigenvalues of
\(V_{\theta_1}^{\mathrm{eff}}(\beta^\star,t)\). Consequently, the same two states remain extremal eigenvectors of the accumulated effective generator, with eigenvalues
\begin{equation}
    \Lambda_{\max/\min}(T)
    =
    \int_0^T
    \lambda_{\max/\min}(t)\,
    \mathrm{d}t.
\end{equation}
It then follows that
\begin{align}
    \left|
        S_{\theta_1}^{\mathrm{eff}}(\beta^\star,T)
    \right|_s
    &=
    \Lambda_{\max}(T)-\Lambda_{\min}(T)
    \nonumber\\
    &=
    \int_0^T
    \left[
        \lambda_{\max}(t)-\lambda_{\min}(t)
    \right]
    \mathrm{d}t
    \nonumber\\
    &=
    \int_0^T
    \left|
        V_{\theta_1}^{\mathrm{eff}}(\beta^\star,t)
    \right|_s
    \mathrm{d}t.
    \label{eq:spectral_bound_saturation}
\end{align}
The control thus saturates the triangle inequality used in deriving Eq.~\eqref{eq:time_independent_qfi_bound}. Moreover, the state in Eq.~\eqref{eq:optimal_initial_state} maximizes the variance of the accumulated effective generator:
\begin{equation}
    4\Delta^2
    \left[
        S_{\theta_1}^{\mathrm{eff}}(\beta^\star,T)
    \right]
    =
    \left|
        S_{\theta_1}^{\mathrm{eff}}(\beta^\star,T)
    \right|_s^2.
    \label{eq:variance_bound_saturation}
\end{equation}
Therefore,
\begin{equation}
    J_{\theta_1\mid\theta_2}^{\star}(T)
    =
    \left[
        \int_0^T
        \left|
            V_{\theta_1}^{\mathrm{eff}}(\beta^\star,t)
        \right|_s
        \mathrm{d}t
    \right]^2,
    \label{eq:ultimate_effective_qfi}
\end{equation}
saturating the upper bound in Eq.(\ref{eq:time_independent_qfi_bound}).

The optimal evolution that saturates the bound is generally not unique. A unitary operator on a \(d\)-dimensional Hilbert space is fully specified by its action on $d$ orthonormal basis, whereas Eq.~\eqref{eq:optimalcontrol} only specifies its action on the two extremal eigenstates. For \(d>2\), the evolution of the remaining \(d-2\) orthogonal directions can be chosen freely. Consequently, there generally exists an infinite family of optimal evolutions, and any evolution that generates a total propagator satisfying Eq.~\eqref{eq:optimalcontrol} achieves the same optimal precision. 

An explicit control Hamiltonian can be constructed from an optimal evolution. Let \(U_{\mathrm{opt}}(t)\) be a unitary evolution that satisfies Eq.~\eqref{eq:optimalcontrol}, the total Hamiltonian that generates it can be obtained as
\begin{equation}
    H_{\rm opt}(t)
    =
    i\dot U_{\rm opt}(t)U^\dagger_{\rm opt}(t).
\end{equation}
Consequently, the optimal control Hamiltonian can then be taken as
\begin{equation}
    H_C(t)
    =
    H_{\star}(t)-H_0(\boldsymbol x,t).
\end{equation}
This control in general depends on the actual value of $\boldsymbol x$. In practice, it can be implemented adaptively by replacing \(\boldsymbol{x}\) with the current estimate \(\hat{\boldsymbol{x}}\) obtained from preliminary or accumulated measurement data. 

An important implication of Eq.(\ref{eq:ultimate_effective_qfi}) is that a nonzero effective QFI can be achieved even when the instantaneous velocity vectors \(V_{\theta_1}(t) = \partial_{\theta_1}H_0(\mathbf{x},t)\) and \(V_{\theta_2}(t) = \partial_{\theta_2}H_0(\mathbf{x},t)\) are collinear at every instant. In such cases, the target parameter \(\theta_1\) might naively appear to be completely indistinguishable from the nuisance parameter \(\theta_2\), since their generators point in the same direction at all times. However, if the ratio between their magnitudes changes with time, the effective generator \(V_{\theta_1}^{\mathrm{eff}}(\beta^\star,t)\) can still acquire a nonzero spectral width. This is because \(\beta^\star\) is a constant chosen to minimize the influence of the nuisance parameter over the entire duration; it cannot compensate for time-dependent variations in the relative strength of the two generators. Thus, even collinear generators can give rise to a nontrivial precision limit for the target parameter, provided their relative magnitudes vary over time. This highlights a key distinction between static and dynamic parameter encoding: time-dependent variations in the generator structure can break the effective degeneracy between parameters, enabling target estimation even when the instantaneous generators are aligned. We illustrate this with some examples in End Matter. 

For time-independent free Hamiltonian $H_0(\boldsymbol x)$, the construction simplifies considerably. In this case, the effective velocity 
\begin{equation}
    V_{\theta_1}^{\rm eff}(\beta^\star,t)
    =
    \partial_{\theta_1}H_0
    -
    \beta^\star\partial_{\theta_2}H_0
\end{equation}
is also time independent, and its extremal eigendirections are fixed, i.e., $|v_{\max/\min}(t)\rangle=|v_{\max/\min}(0)\rangle$. A simple choice of the optimal evolution that satisfies Eq.(\ref{eq:optimalcontrol}) is \(U_{\rm opt}(t)=\mathbb I\). The total Hamiltonian that generates this trivial evolution is $H_{\rm opt}(t)=0$, the corresponding optimal control can thus be taken as \(H_C(t)=H_{\rm opt}(t)-H_0(\hat{x})=-H_0(\hat{x})\). 

This framework extends naturally to the estimation of one target parameter in the presence of multiple nuisance
parameters. Let
\(\boldsymbol\theta_{\rm n}=(\theta_2,\ldots,\theta_N)^T\) denote the nuisance parameters, it can be similarly shown that  
\begin{equation}
    J_{\theta_1|\boldsymbol\theta_{\rm n}}(T)
    =
    4
    \min_{\boldsymbol\beta\in\mathbb R^{N-1}}
    \Delta^2
    \left[
        S_{\theta_1}^{\rm eff}(\boldsymbol\beta,T)
    \right],
    \label{eq:supp_multi_nuisance_projected_generator}
\end{equation}
where $ S_{\theta_1}^{\rm eff}(\boldsymbol\beta,T)
    =
    S_{\theta_1}(T)
    -
    \sum_{\ell=2}^N
    \beta_\ell S_{\theta_\ell}(T)$ is the effective generator with \(\boldsymbol\beta=(\beta_2,\cdots, \beta_N)\).
Following the same procedure as in the two-parameter case, we show that the effective QFI is upper bounded as 
\begin{equation}
    J_{\theta_1|\boldsymbol\theta_{\rm n}}(T)
    \leq
    \left[
    \min_{\boldsymbol\beta\in\mathbb R^{N-1}}
    \int_0^T
    \left|
        V_{\theta_1}^{\rm eff}(\boldsymbol\beta,t)
    \right|_s dt
    \right]^2,
    \label{eq:multi_nuisance_optimal_qfi}
\end{equation}
where 
    \(V_{\theta_1}^{\rm eff}(\boldsymbol\beta,t)
    =
    V_{\theta_1}(t)
    -
    \sum_{\ell=2}^N
    \beta_\ell V_{\theta_\ell}(t)\) with \(V_{j}(t)=\partial_{\theta_j}H_0\).
The minimization over \(\boldsymbol\beta\) is again a convex problem and can be solved similarly as in Eq.(\ref{eq:TD_sdp},\ref{eq:TD_discrete_sdp}).
Once the optimal \(\boldsymbol\beta^\star\) is determined, the optimal probe state and control can be constructed from \(V_{\theta_1}^{\rm eff}(\boldsymbol\beta^\star,t)\), similar to the two-parameter case. In the supplemental material, we provide a detailed description of the optimal protocol, including the optimal probe state, control, and measurement for the estimation of the target parameter in the presence of multiple nuisance parameters. Several examples are also provided in the End Matter to illustrate the general framework.

\paragraph{Summary---}
In this Letter, we have established a general framework for optimal Hamiltonian parameter estimation in the presence of nuisance parameters. By introducing an effective generator that isolates the target signal from variations of the nuisance parameters, we derived the fundamental precision limit for the estimation of the target parameter.  Crucially, we provide an explicit construction of the optimal protocol—encompassing state preparation, control, and measurement-that systematically saturates the fundamental limit. These results bridge a critical gap in multi-parameter quantum metrology and practical sensing applications where not all parameters are known. We anticipate that the effective-generator approach will serve as a versatile template for designing nuisance-robust quantum sensors and for identifying fundamental limits in complex multiparameter estimation tasks. Future directions include generalizing the framework to open quantum systems and multiple target parameters. 

\paragraph{Acknowledgments}--- Z.H. acknowledges Shilin Wang and Allen Zang for the helpful discussions.

We acknowledge support from the Quantum Science and Technology-National Science and Technology Major Project (2023ZD0300600), the Research Grants Council of Hong Kong (14309223, 14309624, 14309022), Guangdong Provincial Quantum Science Strategic Initiative (GDZX2303007,GDZX2505003) and 1+1+1 CUHK-CUHK(SZ)-GDST Joint Collaboration Fund (Grant No. GRDP2025-022), ARO(W911NF-23-1-0077), ARO MURI (W911NF-21-1-0325), AFOSR MURI (FA9550-21-1-0209, FA9550-23-1-0338), ONR MURI (N000142612102), DARPA (HR0011-24-9-0361), NSF (ERC-1941583, OMA-2137642, OSI-2326767, CCF-2312755, OSI-2426975). This work was supported by Quantum Information Science Enabled Discovery 2.0 (QuantISED 2.0) for High Energy Physics (DE-FOA-0003354), with work at Argonne National Laboratory supported by the U.S. Department of Energy under Contract No. DE-AC02-06CH11357.

This material is based upon work supported by the U.S. Department of Energy, Office of Science, National Quantum Information Science Research Centers and Advanced Scientific Computing Research (ASCR) program under contract number DE-AC02-06CH11357 as part of the InterQnet quantum networking project.

\framebox{\parbox{\linewidth}{
The submitted manuscript has been created by UChicago Argonne, LLC, Operator of 
Argonne National Laboratory (``Argonne''). Argonne, a U.S.\ Department of 
Energy Office of Science laboratory, is operated under Contract No.\ 
DE-AC02-06CH11357. 
The U.S.\ Government retains for itself, and others acting on its behalf, a 
paid-up nonexclusive, irrevocable worldwide license in said article to 
reproduce, prepare derivative works, distribute copies to the public, and 
perform publicly and display publicly, by or on behalf of the Government.  The 
Department of Energy will provide public access to these results of federally 
sponsored research in accordance with the DOE Public Access Plan. 
http://energy.gov/downloads/doe-public-access-plan.}}

\bibliography{main_text} 


\clearpage

\section*{End Matter}

We demonstrate the optimal control scheme with two representative examples: estimating a linear frequency drift with an unknown static offset, and estimating the frequency of an AC field with an unknown phase.

\textbf{Example 1: Frequency-drift estimation with an unknown offset.}
Consider a qubit whose transition frequency drifts linearly in time,
\begin{equation}
    H(t)
    =
    \frac{B}{2}
    \left(
        \omega t+\alpha
    \right)
    \sigma_z ,
\end{equation}
where the drift rate \(\omega\) is the target parameter and the static offset 
\(\alpha\) is a nuisance parameter. The corresponding parameter velocities 
are
\begin{equation}
    V_{\omega}(t)
    =
    \frac{B}{2}t\,\sigma_z,
    \qquad
    V_{\alpha}(t)
    =
    \frac{B}{2}\,\sigma_z .
\end{equation}
The effective velocity for the target parameter is
\begin{equation}
    V_{\omega}^{\rm eff}(\beta,t)
    =
    V_{\omega}(t)-\beta V_{\alpha}(t)
    =
    \frac{B}{2}(t-\beta)\sigma_z ,
\end{equation}
which has the spectral width
\(    \left|
        V_{\omega}^{\rm eff}(\beta,t)
    \right|_s
    =
    |B|\,|t-\beta|.\)
The optimal $\beta$, which is given by 
\begin{equation}
\beta^\star=\mathrm{argmin}_{\beta\in\mathbb R}
   \left[\int_0^T \left|
    V_{\theta_1}^{\rm eff}(\beta,t)
    \right|_sdt\right]^2,
    \end{equation}
can be obtained analytically as \(\beta^\star=\frac{T}{2}\), yielding
\begin{equation}
        \min_{\beta\in\mathbb R}
    \int_0^T
    \left|
        V_{\omega}^{\rm eff}(\beta,t)
    \right|_s
    \mathrm{d}t
    =
    \frac{|B|T^2}{4}.
\end{equation}
The optimal effective QFI is therefore
\begin{equation}
    J_{\omega|\alpha}^{\star}(T)
    =
    \frac{B^2T^4}{16}.
    \label{eq:linear_drift_opt_qfi}
\end{equation}

The optimal control to achieve this limit can be realized by a single \(\pi\) pulse about an axis orthogonal to \(z\). This pulse cancels the integrated response to \(\alpha\) while retaining a nonzero response to \({\omega}\). A balanced superposition of the two \(\sigma_z\) eigenstates, followed by the target-specific binary measurement associated with the 
effective generator, attains Eq.~\eqref{eq:linear_drift_opt_qfi} exactly for 
any interrogation time \(T\). 

 This shows that $\omega$ can still be estimated in the presence of an unknown offset $\alpha$, even though both \(V_{\omega}(t)\) and \(V_{\alpha}(t)\) point along the same $\sigma_z$ direction. The key is that the relative magnitude between the two velocity vectors changes with time: \(V_{\omega}(t)\) grows linearly with \(t\), while \(V_{\alpha}(t)\) remains constant. This time-varying proportionality prevents the drift from being absorbed into the static offset, making the two parameters effectively distinguishable despite their collinear generator directions.

\textbf{Example 2: Frequency estimation with an unknown phase.}
Consider a qubit sensor coupled to an AC field,
\begin{equation}
    H(t)
    =
    \frac{B}{2}
    \cos(\omega t+\phi)\sigma_z,
\end{equation}
where \(B\) is known, \(\omega\) is the target frequency, and \(\phi\) is an unknown nuisance phase. The parameter velocities are
\begin{align}
    V_\omega(t)
    &=
    -\frac{B}{2}
    t\sin(\omega t+\phi)\sigma_z,
    \nonumber\\
    V_\phi(t)
    &=
    -\frac{B}{2}
    \sin(\omega t+\phi)\sigma_z.
\end{align}
The effective velocity is therefore
\begin{equation}
    V_\omega^{\rm eff}(\beta,t)
    =
    -\frac{B}{2}
    (t-\beta)\sin(\omega t+\phi)\sigma_z,
\end{equation}
which gives the optimal effecitve QFI
\begin{equation}
    J_{\omega|\phi}^{\star}(T)
    =
    B^2
    \left[
        \int_0^T
        |t-\beta^\star|
        |\sin(\omega t+\phi)|
        \mathrm{d}t
    \right]^2.
    \label{eq:freq_phase_opt_qfi}
\end{equation}
Here \(\beta^\star\) is a weighted median of the interrogation time with weight \(|\sin(\omega t+\phi)|\). This limit can be achieved with the optimal control that consists of \(\pi\)-pulse along an axis orthogonal to $z$ whenever \((t-\beta^\star)\sin(\omega t+\phi)\) changes sign. 
For \(T\gg2\pi/|\omega|\),
\begin{equation}
    \beta^\star
    \simeq
    \frac{T}{2},
    \qquad
    J_{\omega|\phi}^{\star}(T)
    \simeq
    \frac{B^2T^4}{4\pi^2}.
    \label{eq:freq_phase_long_time}
\end{equation}
If the procedure is repeated $\nu$ times, the total effective QFI is then $\simeq
    \frac{\nu B^2T^4}{4\pi^2}$.

For comparison, consider the conventional strategy that directly maximizes the QFI for $\omega$, i.e., maximize $\int_0^T|V_\omega(t)|_sdt$. In this case, the optimal control is to apply $\pi$-pulses along an axis orthogonal to $z$-axis whenever $\sin(\omega t+\phi)$ changes sign. 
The QFIM is given by
\begin{equation}
    J_{\omega}(T)
    =
    B^2
    \begin{pmatrix}
        A_1^2(T)
        &
        A_1(T)A_0(T)
        \\
        A_1(T)A_0(T)
        &
        A_0^2(T)
    \end{pmatrix}.
    \label{eq:freq_phase_single_parameter_qfim}
\end{equation}
where
 \(   A_0(T)
    =
    \int_0^T
    |\sin(\omega t+\phi)|
    \mathrm{d}t,\) and \(
    A_1(T)
    =
    \int_0^T
    t|\sin(\omega t+\phi)|
    \mathrm{d}t.\)
This matrix, however, is singular. A single interrogation duration cannot distinguish \(\omega\) from \(\phi\). The conventional strategy must therefore combine experiments with two different durations.

To get a full rank QFIM for the conventional strategy, we divide the total time, $\nu T$, into two parts. A fraction \(s\) of the total time is used to repeat an evolution of duration \(\tau_1\) with $\frac{s\nu T}{\tau_1}$ times, and the remainder is used to repeat an evolution of duration \(\tau_2\) for $\frac{(1-s)\nu T}{\tau_2}$ times. Let \(J_{\rm conv}(\tau)\) denotes the QFIM of the conventional strategy with the evolution time \(\tau\), the total QFIM is then
\begin{equation}
    J_{\rm conv}^{\rm tot}
    =
    \frac{s\nu T}{\tau_1}
    J_{\rm conv}(\tau_1)
    +
    \frac{(1-s)\nu T}{\tau_2}
    J_{\rm conv}(\tau_2).
\end{equation}
Optimizing $s$, $\tau_1$ and $\tau_2$ leads to
\begin{equation}
    (J_{\rm conv}^{\rm tot})_{\omega|\phi}
    =
    \max_{\substack{
        0<s<1\\
        0<\tau_1,\tau_2\leq T
    }}
    \left[
        (J_{\rm conv}^{\rm tot})_{\omega}
        -
        \frac{
            (J_{\rm conv}^{\rm tot})_{\omega\phi}^2
        }{
            (J_{\rm conv}^{\rm tot})_{\phi}
        }
    \right].
    \label{eq:freq_phase_conventional_information}
\end{equation}

In the long-time regime,
\begin{equation}
    A_0(\tau)
    \simeq
    \frac{2\tau}{\pi},
    \qquad
    A_1(\tau)
    \simeq
    \frac{\tau^2}{\pi},
\end{equation}
and a direct optimization of Eq.~\eqref{eq:freq_phase_conventional_information} gives
\begin{equation}
    \tau_1=T,
    \qquad
    \tau_2=\frac{T}{4},
    \qquad
    s=\frac{1}{3}.
\end{equation}
The resulting effective QFI is
\begin{equation}
    (J_{\rm conv}^{\rm tot})_{\omega|\phi}
    \simeq
    \frac{\nu B^2T^4}{16\pi^2},
    \label{eq:freq_phase_information_comparison}
\end{equation}
Compared to the optimal effective QFI, this is four times smaller.

\section{Adaptive estimates}
The optimal probe, control, and measurement generally depend on the unknown operating point \(\boldsymbol{\theta}_0\). They should
therefore be understood as locally optimal. Practically, this can be achieved with the two-step method \cite{barndorff2000fisher,fujiwara2006strong,gong2026two}: use a small fraction of measurements to obtain an estimation of the operating point first, then implement the optimal protocol based on the estimation. Specifically, a practical implementation can proceed as follows:

\begin{enumerate}
    \item Use an asymptotically negligible fraction of total $\nu$ measurements, for example $\sqrt{\nu}$ measurements, to obtain preliminary estimates
    \(
        \widetilde{\boldsymbol{\theta}}
        =
        (\widetilde{\theta}_1,\widetilde{\theta}_n).
    \)

    \item Evaluate \(
        V_{\theta_1}^{\rm eff}(\boldsymbol\beta,t)
        =
        V_{\theta_1}(t)
        -
        \sum_{j=2}^N\beta_j V_{\theta_j}(t)\) and determine
    \begin{equation}
        \boldsymbol\beta^{\star}
        =
        argmin_\beta
        \int_0^T
        | V_{\theta_1}^{eff}(\boldsymbol\beta,t)|_s\,d t
    \end{equation}
    from the preliminary estimate.

    \item Prepare the optimal probe state, implement the optimal control and perform the optimal measurement according to $\boldsymbol\beta^{\star}$ and the preliminary estimate
    \item Update the estimate of the target parameter.
\end{enumerate}

The nuisance parameters only need to be estimated accurately enough
to localize the operating point and implement the optimal target
protocol. It need not itself be estimated at its ultimate precision. We can periodically use a set of measurements to update estimates of the nuisance estimates when they are needed. The analysis of the robustness of this method can be found in the supplementary material.

\clearpage
\onecolumngrid

\begin{center}
    \mbox{\Large \textbf{Supplemental Material}}

\end{center}

\section{Effective quantum Fisher information with nuisance parameter}
\label{sec:supp_effective_qfi_nuisance}

In this section, we derive the effective quantum Fisher information (QFI)
used in the main text. We first consider estimating the target parameter
\(\theta_1\) at the presence of one nuisance parameter \(\theta_2\).

Under a unitary evolution \(U_{\rm tot}(T)\), the generators of the parameters are given by
\(    S_{\theta_j}(T)
    =
    iU_{\rm tot}^\dagger(T)\partial_{\theta_j}U_{\rm tot}(T),\)
     \(j=1,2\).
For a pure probe state \(\ket{\varphi_0}\), the quantum Fisher information
matrix (QFIM) elements are
\begin{equation}
     [J_{\mathbf \theta}(T)]_{jk}
    =
    4\,\mathrm{Re}
    \left[
    \langle S_{\theta_j}(T)S_{\theta_k}(T)\rangle
    -
    \langle S_{\theta_j}(T)\rangle
    \langle S_{\theta_k}(T)\rangle
    \right],
    \qquad j,k=1,2 ,
    \label{eq:supp_qfim_generator_form}
\end{equation}
where
\(\operatorname{Cov}(A,B)
=
\frac{1}{2}
\langle AB+BA\rangle
-
\langle A\rangle\langle B\rangle.\)
The QFIM can then be written in block form as
\(    J_{\boldsymbol\theta}(T)
    =
    \begin{pmatrix}
        J_{\theta_1}(T)
        &
        J_{\theta_1\theta_2}(T)
        \\
        J_{\theta_2\theta_1}(T)
        &
        J_{\theta_2}(T)
    \end{pmatrix}.\)
When the QFIM is nonsingular, the quantum Cramér-Rao bound (QCRB) for estimating \(\theta_1\) gives
\begin{equation}
    \mathrm{Var}(\hat\theta_1)
    \ge
    \frac{1}{\nu}
    \left[
        J_{\boldsymbol\theta}^{-1}(T)
    \right]_{11}
    =
    \frac{1}{\nu J_{\theta_1|\theta_2}(T)} ,
    \label{eq:supp_effective_qfi_definition}
\end{equation}
where
\(    J_{\theta_1|\theta_2}(T)
    =
    J_{\theta_1}(T)
    -
    \frac{
        J_{\theta_1\theta_2}(T)
        J_{\theta_2\theta_1}(T)
    }{
        J_{\theta_2}(T)
    }\)
is the Schur complement of the nuisance block $J_{\theta_2}(T)$.
When $J_{\theta_2}(T)$ is singular, the effective QFI is defined through
the generalized Schur complement,
\begin{equation}
    J_{\theta_1|\theta_2}(T)
    =
    J_{\theta_1}(T)
    -
    J_{\theta_1\theta_2}(T)
    J_{\theta_2}^{+}(T)
    J_{\theta_2\theta_1}(T),
    \label{eq:supp_effective_qfi_generalized_schur}
\end{equation}
where \(J_{\theta_2\theta_2}^{+}(T)\) denotes the Moore-Penrose
pseudoinverse. In the extreme case where
\(J_{\theta_2}(T)=0\), positivity of the QFIM implies that the off-diagonal block must also vanish,
\(J_{\theta_1\theta_2}(T)=0\). Consequently, the effective QFI reduces simply to
\(    J_{\theta_1|\theta_2}(T)
    =
    J_{\theta_1}(T).\)

We now seek a more transparent representation in terms of the generators. Using Eq.~\eqref{eq:supp_qfim_generator_form}, for
the nonsingular case \(\Delta^2S_{\theta_2}(T)>0\), the Schur complement can
be rewritten as
\begin{align}
    J_{\theta_1|\theta_2}(T)
    &=
    4
    \left[
    \Delta^2 S_{\theta_1}(T)
    -
    \frac{
    \mathrm{Cov}^2
    \left(
        S_{\theta_1}(T),S_{\theta_2}(T)
    \right)
    }{
    \Delta^2 S_{\theta_2}(T)
    }
    \right] .
    \label{eq:supp_schur_generator_covariance}
\end{align}
We now introduce the effective generator
\(    S_{\theta_1}^{\rm eff}(\beta,T)
    =
    S_{\theta_1}(T)-\beta S_{\theta_2}(T)\) with a tunable weight $\beta\in \mathbb{R}$.
Its variance is given by
\begin{align}
    \Delta^2
    S_{\theta_1}^{\rm eff}(\beta,T)
    &=
    \Delta^2S_{\theta_1}(T)
    -
    2\beta
    \left[
    \mathrm{Cov}
    \left(
        S_{\theta_1}(T),S_{\theta_2}(T)
    \right)
    \right]
    +
    \beta^2
    \Delta^2S_{\theta_2}(T).
    \label{eq:supp_effective_generator_variance}
\end{align}

When \(\Delta^2S_{\theta_2}(T)>0\), the optimal $\beta$ that minimizes this variance is given by
\begin{equation}
    \beta^{\star}
    =
    \frac{
    \mathrm{Cov}
    \left(
        S_{\theta_1}(T),S_{\theta_2}(T)
    \right)
    }{
    \Delta^2S_{\theta_2}(T)
    } .
    \label{eq:supp_beta^probe_dependent}
\end{equation}
And the resulting minimal $\Delta^2
    S_{\theta_1}^{\rm eff}(\beta^\star,T)$
equals $J_{\theta_1|\theta_2}(T)$ in
Eq.~\eqref{eq:supp_schur_generator_covariance}.
For the singular case
\(\Delta^2S(T)=0\), positivity of the
covariance matrix forces
\(    \mathrm{Cov}
    \left(
        S_{\theta_1}(T),S_{\theta_2}(T)
    \right)=    0.\)
So the variance in
Eq.~\eqref{eq:supp_effective_generator_variance} becomes independent of
\(\beta\). Thus, the minimization over $\beta$
is trivial and yields the same effective QFI.

Combining both cases, we obtain the unified variational representation
\begin{equation}
    J_{\theta_1|\theta_2}(T)
    =
    4
    \min_{\beta\in\mathbb R}
    \Delta^2
    \left[
        S_{\theta_1}^{\rm eff}(\beta,T)
    \right].
    \label{eq:supp_effective_qfi_projected_generator}
\end{equation}

\section{Choice of nuisance parameter}
We show that the particular choice of the nuisance coordinate used in the main text is merely a matter of convenience and does not affect the effective QFI. 

Suppose the original parameters are \((x_1,x_2)\), and the target parameter of interest is
\begin{equation}
    \theta_1
    =
    w_1x_1+w_2x_2,
    \qquad
    w_1^2+w_2^2=1 .
    \label{eq:supp_target_coordinate}
\end{equation}
In the main text, we choose the nuisance parameter to be the orthogonal combination
\begin{equation}
    \theta_2
    =
    w_2x_1-w_1x_2 .
    \label{eq:supp_orthogonal_nuisance_coordinate}
\end{equation}
With this choice, the corresponding generators are given by
\begin{equation}
    S_{\theta_1}(T)
    =
    w_1S_{x_1}(T)+w_2S_{x_2}(T),
    \qquad
    S_{\theta_2}(T)
    =
    w_2S_{x_1}(T)-w_1S_{x_2}(T).
    \label{eq:supp_orthogonal_generators}
\end{equation}

We now show that the effective QFI is independent of this choice. To this end, consider a general nuisance parameter
\(    \theta_2'
    =
    v_1x_1+v_2x_2\)
that is linearly independent of \(\theta_1\), meaning
\begin{equation}
   D
    :=
    w_1v_2-w_2v_1
    \neq0 .
   \label{eq:supp_coordinate_det}
\end{equation}
The coordinate transformation from \((x_1,x_2)\) to \((\theta_1,\theta_2')\) is
\begin{equation}
    \begin{pmatrix}
        \theta_1\\
        \theta_2'
    \end{pmatrix}
    =
    \begin{pmatrix}
        w_1 & w_2\\
        v_1 & v_2
    \end{pmatrix}
    \begin{pmatrix}
        x_1\\
        x_2
    \end{pmatrix},
\end{equation}
which has the inverse
\begin{equation}
    \begin{pmatrix}
        x_1\\
        x_2
    \end{pmatrix}
    =
    \frac{1}{D}
    \begin{pmatrix}
        v_2 & -w_2\\
        -v_1 & w_1
    \end{pmatrix}
    \begin{pmatrix}
        \theta_1\\
        \theta_2'
    \end{pmatrix}.
    \label{eq:supp_general_coordinate_inverse}
\end{equation}
The generators for \((\theta_1,\theta_2')\) are then given by
\begin{align}
    S_{\theta_1}'(T)
    &=
    \frac{\partial x_1}{\partial\theta_1}
    S_{x_1}(T)
    +
    \frac{\partial x_2}{\partial\theta_1}
    S_{x_2}(T)=
    \frac{v_2}{D}S_{x_1}(T)
    -
    \frac{v_1}{D}S_{x_2}(T),
    \label{eq:supp_general_target_generator}
    \\
    S_{\theta_2'}(T)
    &=
    \frac{\partial x_1}{\partial\theta_2'}
    S_{x_1}(T)
    +
    \frac{\partial x_2}{\partial\theta_2'}
    S_{x_2}(T)=
    -\frac{w_2}{D}S_{x_1}(T)
    +
    \frac{w_1}{D}S_{x_2}(T).
    \label{eq:supp_general_nuisance_generator}
\end{align}

To relate this to the orthogonal coordinate choice, we first express the original variables in terms of \((\theta_1,\theta_2)\):
\[
x_1 = w_1\theta_1 + w_2\theta_2, \qquad
x_2 = w_2\theta_1 - w_1\theta_2.
\]
We can then write the new nuisance parameter  $\theta_2'$ in terms of the orthogonal choice \((\theta_1,\theta_2)\) as
\begin{align}
    \theta_2'=
    v_1x_1+v_2x_2=
    (v_1w_1+v_2w_2)\theta_1
    +
    (v_1w_2-v_2w_1)\theta_2=
    a\theta_1-D\theta_2 ,
    \label{eq:supp_theta2_prime_aD}
\end{align}
where we have defined
\(    a=v_1w_1+v_2w_2 .\)
Equivalently, we can invert this relation to obtain
\begin{equation}
    \theta_2
    =
    \frac{a\theta_1-\theta_2'}{D},
    \label{eq:supp_theta2_inverse_general}
\end{equation}
therefore,
\begin{equation}
    \begin{pmatrix}
        \theta_1\\
        \theta_2
    \end{pmatrix}
    =
    \begin{pmatrix}
        1 & 0\\
        \frac{a}{D} & -\frac{1}{D}
    \end{pmatrix}
    \begin{pmatrix}
        \theta_1\\
        \theta_2'
    \end{pmatrix}.
\end{equation}
Using the chain rule, we then have
\begin{eqnarray}
    \aligned
    S_{\theta_1}'(T)
    &=\frac{\partial \theta_1}{\partial \theta_1}S_{\theta_1}(T)
    +
    \frac{\partial \theta_2}{\partial \theta_1}S_{\theta_2}(T)=
    S_{\theta_1}(T)
    +
    \frac{a}{D}S_{\theta_2}(T),
    \\
    S_{\theta_2'}(T)
    &=\frac{\partial \theta_1}{\partial \theta_2'}S_{\theta_1}(T)
    +
    \frac{\partial \theta_2}{\partial \theta_2'}S_{\theta_2}(T)=
    -\frac{1}{D}S_{\theta_2}(T).
   \endaligned 
\end{eqnarray}
Thus, changing the nuisance coordinate has two effects: it shifts the target generator \(S_{\theta_1}(T)\) by a term proportional to the orthogonal nuisance generator \(S_{\theta_2}(T)\), and it rescales the nuisance generator itself.

We now verify that the effective QFI is invariant under this transformation. Using the variational representation from Eq.~(\ref{eq:supp_effective_qfi_projected_generator}), we have
\begin{align}
    J_{\theta_1|\theta_2'}(T)
    &=
    4
    \min_{\beta'\in\mathbb R}
    \Delta^2
    \left[
        S_{\theta_1}'(T)-\beta' S_{\theta_2'}(T)
    \right]
    \nonumber\\
    &=
    4
    \min_{\beta'\in\mathbb R}
    \Delta^2
    \left[
        S_{\theta_1}(T)
        +
        \frac{a}{D}S_{\theta_2}(T)
        +
        \frac{\beta'}{D}S_{\theta_2}(T)
    \right]
    \nonumber\\
    &=
    4
    \min_{\gamma\in\mathbb R}
    \Delta^2
    \left[
        S_{\theta_1}(T)-\gamma S_{\theta_2}(T)
    \right]
    \nonumber\\
    &=
    J_{\theta_1|\theta_2}(T),
    \label{eq:supp_coordinate_invariance_explicit}
\end{align}
where we have defined
$\gamma
    =
    -\frac{a+\beta'}{D}.$
Since \(D\neq0\), the mapping \(\beta' \mapsto \gamma\) is a bijection from \(\mathbb R\) onto itself. Therefore, the minimization over \(\beta'\) is equivalent to the minimization over \(\gamma\), and the effective QFI remains unchanged. 

This establishes that the effective QFI is independent of the particular nuisance coordinate chosen to complete the target parameter. The orthogonal choice \(\theta_2 = w_2x_1 - w_1x_2\) is therefore only a convenient convention.

\section{Consistency of the optimal nuisance coefficient}
\label{sec:supp_extremal_probe_condition}

In the main text, the optimal nuisance coefficient \(\beta^\star\) is defined in two different ways:

1. As the minimizer of the integrated spectral norm of the instantaneous effective generator:
   \begin{equation}\label{eq:beta1}
   \beta^\star
   =
   \operatorname*{arg\,min}_{\beta\in\mathbb R}
   \int_0^T
   \left|
   V_{\theta_1}^{\rm eff}(\beta,t)
   \right|_s
   \mathrm{d}t.
   \end{equation}

2. As the minimizer of the variance of the accumulated effective generator:
   \begin{equation}\label{eq:beta2}
   \beta^\star
   =
   \operatorname*{arg\,min}_{\beta\in\mathbb R}
   4\Delta^2 S_{\theta_1}^{\rm eff}(\beta,T).
   \end{equation}

We now show that these two definitions are in fact consistent: the \(\beta^\star\) that minimizes the integrated instantaneous spectral norm also minimizes the accumulated variance, provided that the probe state is chosen optimally.

Recall the definitions of the effective generators:
\begin{equation}
    S_{\theta_1}^{\rm eff}(\beta,T)
    =
    S_{\theta_1}(T)-\beta S_{\theta_2}(T),
\end{equation}
their instantaneous counterparts
\(    V_{\theta_1}^{\rm eff}(\beta,t)
    =
    V_{\theta_1}(t)-\beta V_{\theta_2}(t).\)
Let \(\beta^\star
=
\operatorname*{arg\,min}_{\beta\in\mathbb R}
\int_0^T
\left|
V_{\theta_1}^{\rm eff}(\beta,t)
\right|_s
\mathrm{d}t\). Denote by \(\ket{v_{\max}(t)}\) and \(\ket{v_{\min}(t)}\) the eigenstates corresponding to the largest and smallest eigenvalues of \(V_{\theta_1}^{\rm eff}(\beta^\star,t)\), respectively. We now consider the optimal probe state constructed from these extremal eigenvectors at the initial time: 
 \(\ket{\varphi_0^\star}
    =
    \frac{
        \ket{v_{\max}(0)}
        +
        e^{i\chi}\ket{v_{\min}(0)}
    }{\sqrt{2}}.\) We will prove that for this probe, \(\beta^\star\) indeed minimizes \(\Delta^2 S_{\theta_1}^{\rm eff}(\beta,T)\). 

First, when the largest and smallest eigenvalues of
\(V_{\theta_1}^{\rm eff}(\beta^\star,t)\) are nondegenerate, denote
these eigenvalues by \(\lambda_{\max}(t)\) and \(\lambda_{\min}(t)\), with
corresponding eigenstates \(\ket{v_{\max}(t)}\) and
\(\ket{v_{\min}(t)}\). The Hellmann--Feynman theorem gives
\begin{align}
    \left.
    \frac{\mathrm{d}}{\mathrm{d}\beta}
    \int_0^T
    \left|
        V_{\theta_1}^{\rm eff}(\beta,t)
    \right|_s
    \mathrm{d}t
    \right|_{\beta=\beta^\star}
=
    -
    \int_0^T
    \bigg[
        \bra{v_{\max}(t)}
        V_{\theta_2}(t)
        \ket{v_{\max}(t)}
        -
        \bra{v_{\min}(t)}
        V_{\theta_2}(t)
        \ket{v_{\min}(t)}
    \bigg]
    \mathrm{d}t=0 .\label{eq:supp_integrated_nuisance_balance}
\end{align}
Now consider the control strategy introduced in the main text, which keeps the extremal eigenvectors aligned throughout the evolution:
 \(   U_{\rm tot}(t)
    \ket{v_{\max/\min}(0)}
    =
    \ket{v_{\max/\min}(t)}.\)
Under this control, the accumulated effective generator acts on the initial extremal states as
\begin{equation}
    S_{\theta_1}^{\rm eff}(\beta^\star,T)
    \ket{v_{\max/\min}(0)}
    =
    \Lambda_{\max/\min}(T)
    \ket{v_{\max/\min}(0)},
\end{equation}
where the accumulated eigenvalues are given by
 \(   \Lambda_{\max/\min}(T)
    =
    \int_0^T
    \lambda_{\max/\min}(t)
    \mathrm{d}t.\)
We then have
\begin{align}
    \left|
        S_{\theta_1}^{\rm eff}(\beta^\star,T)
    \right|_s
    =
    \Lambda_{\max}(T)-\Lambda_{\min}(T)
    =
    \int_0^T
    \left|
        V_{\theta_1}^{\rm eff}(\beta^\star,t)
    \right|_s
    \mathrm{d}t .
    \label{eq:supp_accumulated_width_saturation}
\end{align}

Next, we examine the nuisance responses accumulated along the two extremal branches. These are given by
\begin{align}
    \bra{v_{\max/\min}(0)}
    S_{\theta_2}(T)
    \ket{v_{\max/\min}(0)}
    =
    \int_0^T
    \bra{v_{\max/\min}(t)}
    V_{\theta_2}(t)
    \ket{v_{\max/\min}(t)}
    \mathrm{d}t.
\end{align}
From the condition in Eq.~\eqref{eq:supp_integrated_nuisance_balance},we then have
\begin{equation}
    \bra{v_{\max}(0)}
    S_{\theta_2}(T)
    \ket{v_{\max}(0)}
    =
    \bra{v_{\min}(0)}
    S_{\theta_2}(T)
    \ket{v_{\min}(0)} .
    \label{eq:supp_equal_nuisance_response}
\end{equation}

Now consider the optimal probe
 \(   \ket{\varphi_0^\star}
    =
    \frac{
        \ket{v_{\max}(0)}
        +
        e^{i\chi}\ket{v_{\min}(0)}
    }{\sqrt{2}}.\)
For this probe state, the variance of the accumulated effective generator saturates the spectral-width variance bound:
\begin{align}
    4
    \Delta^2
    S_{\theta_1}^{\rm eff}(\beta^\star,T)
    =
    \left|
        S_{\theta_1}^{\rm eff}(\beta^\star,T)
    \right|_s^2
=
    \left[
        \int_0^T
        \left|
            V_{\theta_1}^{\rm eff}(\beta^\star,t)
        \right|_s
        \mathrm{d}t
    \right]^2 .
    \label{eq:supp_width_saturation}
\end{align}
Furthermore, using Eq.~\eqref{eq:supp_equal_nuisance_response}, we find that the covariance between the effective generator and the nuisance generator vanishes:
\begin{align}
        \mathrm{Cov}
        \left(
            S_{\theta_1}^{\rm eff}(\beta^\star,T),
            S_{\theta_2}(T)
        \right)
    =
    \frac{
        \Lambda_{\max}(T)-\Lambda_{\min}(T)
    }{4}
    \bigg[
        \bra{v_{\max}(0)}
        S_{\theta_2}(T)
        \ket{v_{\max}(0)}
        -
        \bra{v_{\min}(0)}
        S_{\theta_2}(T)
        \ket{v_{\min}(0)}
    \bigg]
    =0.
    \label{eq:supp_cov_zero}
\end{align}

We now verify that \(\beta^\star\) also minimizes \[\Delta^2
    \left[
        S_{\theta_1}(T)
        -
        \beta S_{\theta_2}(T)
    \right]=\langle[\int_0^TV_{\theta_1}^{\rm eff}(t)dt]^2\rangle-\langle\int_0^TV_{\theta_1}^{\rm eff}(t)dt\rangle^2. \]
To show this, let    
\(\tilde\beta=\beta^\star+\delta\beta\), we then have
\begin{align}
    \Delta^2
    \left[
        S_{\theta_1}(T)
        -
        \tilde\beta S_{\theta_2}(T)
    \right]
    =
    \Delta^2
    S_{\theta_1}^{\rm eff}(\beta^\star,T)
    -
    2\delta\beta\,
        \mathrm{Cov}
        \left(
            S_{\theta_1}^{\rm eff}(\beta^\star,T),
            S_{\theta_2}(T)
        \right)
    +
    (\delta\beta)^2
    \Delta^2
    S_{\theta_2}(T).
\end{align}
Using Eq.~\eqref{eq:supp_cov_zero}, this simplifies to
\begin{align}
    \Delta^2
    \left[
        S_{\theta_1}(T)
        -
        \tilde\beta S_{\theta_2}(T)
    \right]
    =
    \Delta^2
    S_{\theta_1}^{\rm eff}(\beta^\star,T)
    +
    (\delta\beta)^2
    \Delta^2
    S_{\theta_2}(T)
   \geq
    \Delta^2
    S_{\theta_1}^{\rm eff}(\beta^\star,T).
\end{align}
Therefore,
\begin{equation}
    \min_{\tilde\beta\in\mathbb R}
    \Delta^2
    \left[
        S_{\theta_1}(T)
        -
        \tilde\beta S_{\theta_2}(T)
    \right]
    =
    \Delta^2
    S_{\theta_1}^{\rm eff}(\beta^\star,T).
    \label{eq:supp_min_attained_beta^star}
\end{equation}
This proves that \(\beta^\star\) from Eq. (\ref{eq:beta1}) also satisfies Eq. (\ref{eq:beta2}) when evaluated on the optimal probe state. The same construction remains valid when the extremal eigenvalues are degenerate; in that case, the extremal states should be chosen within the corresponding eigenspaces to satisfy the balance condition in Eq. (\ref{eq:supp_equal_nuisance_response}).

Thus, the two definitions of \(\beta^\star\) are consistent: the nuisance coefficient obtained from minimizing the integrated spectral norm of the instantaneous generator also minimizes the variance of the accumulated generator for the optimal probe. This consistency ensures that our protocol is self-contained and that the optimization over the control and the minimization over the nuisance coefficient can be performed in either order without affecting the final result.

\section{Optimal scheme}
\label{sec:supp_optimal_scheme}

In this section, we present the full optimal protocol, including the optimal probe state, control, and measurement, for estimating the target parameter \(\theta_1\) in the presence of a nuisance parameter \(\theta_2\). This shows explicitly that the upper bound derived earlier is indeed attainable.

Let \(\beta^\star=\operatorname*{arg\,min}_{\beta\in\mathbb R}
\int_0^T
\left|
V_{\theta_1}^{\rm eff}(\beta,t)
\right|_s
\mathrm{d}t\), \(\ket{v_{\max}(t)}\) and
\(\ket{v_{\min}(t)}\) as the eigenstates associated with the largest
and smallest eigenvalues of
\(V_{\theta_1}^{\rm eff}(\beta^\star,t)\), respectively. The optimal initial probe state is taken as
\begin{equation}
    \ket{\varphi_0^\star}
    =
    \frac{
        \ket{v_{\max}(0)}
        +
        \ket{v_{\min}(0)}
    }{\sqrt{2}}.
    \label{eq:supp_direct_optimal_probe}
\end{equation}

The total evolution under the optimal control transports the two instantaneous extremal
eigenstates according to
\begin{equation}
    U_{\rm tot}(\boldsymbol\theta_0,t)
    \ket{v_{\max/\min}(0)}
    =
    \ket{v_{\max/\min}(t)},
    \qquad
    0\leq t\leq T,
    \label{eq:supp_extremal_state_transport}
\end{equation}
where $\theta_0$ is the operating point. 
Consequently, \(\ket{v_{\max}(0)}\) and \(\ket{v_{\min}(0)}\) remain the extreme eigenstates of the accumulated effective generator \(S_{\theta_1}^{\rm eff}(\beta^\star,T)=
    \int_0^T
    U_{\rm tot}^\dagger(t)
    V_{\theta_1}^{\rm eff}(\beta,t)
    U_{\rm tot}(t)\,dt\). For this optimal initial state, the variance of the effective generator saturates the spectral-width variance bound:
\(    4
    \Delta^2
    S_{\theta_1}^{\rm eff}(\beta^\star,T)
    =
    \left|
        S_{\theta_1}^{\rm eff}(\beta^\star,T)
    \right|_s^2\).
Furthermore, under the optimal evolution, the spectral width of the accumulated generator equals the integrated instantaneous width:
\begin{equation}
    \left|
        S_{\theta_1}^{\rm eff}(\beta^\star,T)
    \right|_s
    =
    \int_0^T
    \left|
        V_{\theta_1}^{\rm eff}(\beta^\star,t)
    \right|_s
    \mathrm{d}t .
\end{equation}

The QFIM, with its entries given by 
$    \left[
        J_{\boldsymbol\theta}(T)
    \right]_{jk}
    =
    4 {\rm Re}
    \left[
        \mathrm{Cov}
        \left(
            S_{\theta_j}(T),
            S_{\theta_k}(T)
        \right)
    \right]$,
is
\begin{equation}
    J_{\boldsymbol\theta}^{\star}(T)
    =
    \begin{pmatrix}
        \left|
            S_{\theta_1}^{\rm eff}(\beta^\star,T)
        \right|_s^2
        +
        \beta^{\star^2}J_{\theta_2}(T)
        &
        \beta^\star J_{\theta_2}(T)
        \\[1mm]
        \beta^\star J_{\theta_2}(T)
        &
        J_{\theta_2}(T)
    \end{pmatrix},
    \label{eq:supp_optimal_qfim_direct}
\end{equation}
where we used the fact that $S_{\theta_1}(T)
=
S_{\theta_1}^{\rm eff}(\beta^\star,T)
+
\beta^\star S_{\theta_2}(T)$ and $\operatorname{Cov}\left(
S_{\theta_1}^{\rm eff}(\beta^\star,T),
S_{\theta_2}(T)
\right)
=
0$, here
 $   J_{\theta_2}(T)
    =
    4
    \Delta^2
    S_{\theta_2}(T).$
Taking the Schur complement of the nuisance block yields
\begin{equation}
    J_{\theta_1|\theta_2}^{\star}(T)
    =
    \left|
        S_{\theta_1}^{\rm eff}(\beta^\star,T)
    \right|_s^2.
    \label{eq:supp_optimal_effective_qfi_direct}
\end{equation}
We note that for the case \(J_{\theta_2}(T)=0\), the same result follows by using the Moore–Penrose pseudoinverse. Thus, the probe and control indeed attain the optimized effective QFI derived in the preceding section.

We now construct a measurement that achieves this bound. Define the observable
\begin{equation}
    O_{\theta_1}^{\rm eff}
    =
    -i
    \ket{v_{\max}(0)}
    \bra{v_{\min}(0)}
    +
    i
    \ket{v_{\min}(0)}
    \bra{v_{\max}(0)}.
    \label{eq:supp_optimal_measurement_observable}
\end{equation}
For the optimal probe, this observable satisfies the relation
\begin{equation}
    \left[
        S_{\theta_1}^{\rm eff}(\beta^\star,T)
        -
        \left\langle
            S_{\theta_1}^{\rm eff}(\beta^\star,T)
        \right\rangle
    \right]
    \ket{\varphi_0^\star}
    =
    i
    \frac{
        \left|
            S_{\theta_1}^{\rm eff}(\beta^\star,T)
        \right|_s
    }{2}
    O_{\theta_1}^{\rm eff}
    \ket{\varphi_0^\star}.
    \label{eq:supp_optimal_measurement_condition}
\end{equation}
Hence, the eigenbasis of \(O_{\theta_1}^{\rm eff}\) forms the optimal measurement \cite{hou2021super}. This effective optimal observable is defined in the Heisenberg picture, same as the effective generator. In the Schrodinger picture, the optimal observable is given by $U_{tot}(\theta_0, T)O_{\theta_1}^{\rm eff}U_{tot}(\theta_0, T)$ with the corresponding measurement basis given by
\begin{equation}
    \ket{\pi_\pm}
    =
    U_{\rm tot}(\boldsymbol\theta_0,T)
    \frac{
        \ket{v_{\max}(0)}
        \pm
        i\ket{v_{\min}(0)}
    }{\sqrt{2}}
    =
    \frac{
        \ket{v_{\max}(T)}
        \pm
        i\ket{v_{\min}(T)}
    }{\sqrt{2}}.
    \label{eq:supp_optimal_measurement}
\end{equation}
To form a complete POVM in a larger Hilbert space, it can be written as
\begin{equation}
M_\pm
    =
    \ket{\pi_\pm}\bra{\pi_\pm}
    +
    \frac{1}{2}
    \left(
        \mathbb I
        -
        \ket{\pi_+}\bra{\pi_+}
        -
        \ket{\pi_-}\bra{\pi_-}
    \right)
    \label{eq:supp_optimal_measurement_povm}
\end{equation}
which satisfies \(M_++M_-=\mathbb I\).

Since the output state is $\ket{\varphi_T^\star}
=
U_{\rm tot}(\boldsymbol\theta_0,T)\ket{\varphi_0^\star}
=
\frac{
\ket{v_{\max}(T)} + \ket{v_{\min}(T)}
}{\sqrt{2}}$,
the outcome probabilities,
$    p_\pm(\boldsymbol\theta)
    =
    \bra{\varphi_T^\star}
    M_{\pm}
    \ket{\varphi_T^\star},$
are \(p_\pm(\boldsymbol\theta_0)=1/2\). 
To compute the classical Fisher information, we first take the derivatives of the output state,
\[
\partial_{\theta_j}\ket{\psi_T^\star(\boldsymbol\theta_0)}
=
-iU(\theta, T)S_{\theta_j}(T)U^\dagger(\theta, T)\ket{\psi_T^\star(\boldsymbol\theta_0)},
\qquad j=1,2,
\]
here $U(\theta, T)S_{\theta_j}(T)U^\dagger(\theta, T)$ is the generator of $\theta_j$ in the Schrodinger picture. 
For the target parameter, we use the relation \(S_{\theta_1}(T) = S_{\theta_1}^{\rm eff}(\beta^\star,T) + \beta^\star S_{\theta_2}(T)\), so that
\begin{equation}\label{eq:derivative1}
\partial_{\theta_1}\ket{\psi_T^\star(\boldsymbol\theta_0)}
=
-iU(\theta,T)\left[S_{\theta_1}^{\rm eff}(\beta^\star,T) + \beta^\star S_{\theta_2}(T)\right]U^\dagger(\theta,T)\ket{\psi_T^\star(\boldsymbol\theta_0)}.
\end{equation}
For the nuisance parameter, we simply have
\begin{equation}\label{eq:derivative2}
\partial_{\theta_2}\ket{\psi_T^\star(\boldsymbol\theta_0)}
=
-iU(\theta,T)S_{\theta_2}(T)U\dagger(\theta,T)\ket{\psi_T^\star(\boldsymbol\theta_0)}.
\end{equation}
We now compute the derivatives of the probability difference \(\Delta p := p_+ - p_-\). Since \(M_+ - M_- = \ket{\pi_+}\bra{\pi_+} - \ket{\pi_-}\bra{\pi_-}\), we have
\begin{equation}\label{eq:derivativep}
\Delta p(\boldsymbol\theta)
=
\bra{\psi_T^\star(\boldsymbol\theta)}
\left(M_+ - M_-\right)
\ket{\psi_T^\star(\boldsymbol\theta)}.
\end{equation}
At the operating point, the derivative with respect to \(\theta_j\) is
\begin{equation}\label{eq:derivativepsupp}
\left.
\partial_{\theta_j}
\Delta p
\right|_{\boldsymbol\theta_0}
=
2\,\mathrm{Re}\left[
\bra{\psi_T^\star(\boldsymbol\theta_0)}
\left(M_+ - M_-\right)
\partial_{\theta_j}\ket{\psi_T^\star(\boldsymbol\theta_0)}
\right],
\end{equation}
Substituting Eqs. (\ref{eq:derivative2}) and (\ref{eq:derivative2}) into Eq. (\ref{eq:derivativepsupp}), and using the explicit form of the measurement projectors, we obtain
\[
\left.
\partial_{\theta_j}
\Delta p
\right|_{\boldsymbol\theta_0}
=
\frac{4}{
\left|
S_{\theta_1}^{\rm eff}(\beta^\star,T)
\right|_s
}
\operatorname{Cov}\left(
S_{\theta_1}^{\rm eff}(\beta^\star,T),
S_{\theta_j}(T)
\right),
\qquad j=1,2.
\]
For the nuisance parameter \(j=2\), the optimality condition derived in the preceding section ensures that the covariance vanishes:
\(
\operatorname{Cov}\left(
S_{\theta_1}^{\rm eff}(\beta^\star,T),
S_{\theta_2}(T)
\right)
=
0.
\)
Therefore,
\(
\left.
\partial_{\theta_2}
\Delta p
\right|_{\boldsymbol\theta_0}
=
0.
\)
For the target parameter \(j=1\), we evaluate the covariance:
\[
\operatorname{Cov}\left(
S_{\theta_1}^{\rm eff}(\beta^\star,T),
S_{\theta_1}(T)
\right)
=
\operatorname{Cov}\left[
S_{\theta_1}^{\rm eff}(\beta^\star,T),
S_{\theta_1}^{\rm eff}(\beta^\star,T) + \beta^\star S_{\theta_2}(T)
\right]=\Delta^2 S_{\theta_1}^{\rm eff}(\beta^\star,T).
\]
Together with the condition
\(
4\Delta^2 S_{\theta_1}^{\rm eff}(\beta^\star,T)
=
\left|
S_{\theta_1}^{\rm eff}(\beta^\star,T)
\right|_s^2,
\)
we obtain
\begin{equation}
\left.
\partial_{\theta_1}
\Delta p
\right|_{\boldsymbol\theta_0}
=
\left|
S_{\theta_1}^{\rm eff}(\beta^\star,T)
\right|_s.
\end{equation}
Thus, to first order in the parameter deviations,
\begin{equation}
p_\pm(\boldsymbol\theta_0 + \delta\boldsymbol\theta)
=
\frac{1}{2}
\pm
\frac{1}{2}
\left|
S_{\theta_1}^{\rm eff}(\beta^\star,T)
\right|_s
\delta\theta_1
+
O\!\left(
\|\delta\boldsymbol\theta\|^2
\right).
\end{equation}
Crucially, the nuisance parameter produces no first-order contribution to either outcome probability. The measurement therefore effectively eliminates the nuisance sensitivity while retaining full information about the target parameter.

Since \(p_\pm(\boldsymbol\theta_0) = 1/2\) and the derivatives of the individual probabilities satisfy \(\partial_{\theta_j}p_+ = -\partial_{\theta_j}p_- = \frac{1}{2}\partial_{\theta_j}\Delta p\), we have
\[
F(\boldsymbol\theta_0,T)
=\left.
\sum_{\alpha=\pm}
\frac{
\partial_{\theta_i}p_\alpha(\boldsymbol\theta)
\,
\partial_{\theta_j}p_\alpha(\boldsymbol\theta)
}{
p_\alpha(\boldsymbol\theta)
}
\right|_{\boldsymbol\theta=\boldsymbol\theta_0}=
\begin{pmatrix}
\left(\partial_{\theta_1}\Delta p\right)^2 & \left(\partial_{\theta_1}\Delta p\right)\left(\partial_{\theta_2}\Delta p\right) \\
\left(\partial_{\theta_2}\Delta p\right)\left(\partial_{\theta_1}\Delta p\right) & \left(\partial_{\theta_2}\Delta p\right)^2
\end{pmatrix}=
\begin{pmatrix}
\left|
S_{\theta_1}^{\rm eff}(\beta^\star,T)
\right|_s^2
&
0
\\
0
&
0
\end{pmatrix}..
\]
Although the nuisance block is singular, the cross block also vanishes. Using the generalized Schur complement with the Moore–Penrose pseudoinverse for the nuisance block, we obtain
\begin{equation}
F_{\theta_1|\theta_2}(T)
=
J_{\theta_1|\theta_2}^{\star}(T)
=
\left|
S_{\theta_1}^{\rm eff}(\beta^\star,T)
\right|_s^2.
\end{equation}
Thus, the measurement removes the nuisance response at first order while retaining all effective information about the target parameter. For \(\nu\) independent repetitions and any estimator that is locally unbiased for \(\theta_1\) at \(\boldsymbol\theta_0\) (including along the nuisance direction), the quantum Cramér–Rao bound yields
\begin{equation}
\operatorname{Var}\left(\widehat{\theta}_1\right)
\geq
\frac{1}{
\nu
\left|
S_{\theta_1}^{\rm eff}(\beta^\star,T)
\right|_s^2
}.
\end{equation}

In summary, the probe state, control protocol, and measurement scheme together attain the optimized local precision for the target parameter in the presence of one nuisance parameter. This completes the constructive proof that the upper bound derived earlier is indeed achievable.

\section{Extension to multiple nuisance parameters}

We now extend the derivation to the estimation of one target parameter \(\theta_1\) in the presence of multiple nuisance parameters \(\boldsymbol\theta_{\rm n}=(\theta_2,\ldots,\theta_N)^T\). The QFIM can be written in block form as
\begin{equation}
    J_{\boldsymbol\theta}(T)
    =
    \begin{pmatrix}
        J_{\theta_1}(T)
        &
        J_{\theta_1\boldsymbol\theta_{\rm n}}(T)
        \\
        J_{\boldsymbol\theta_{\rm n}\theta_1}(T)
        &
        J_{\boldsymbol\theta_{\rm n}}(T)
    \end{pmatrix}.
    \label{eq:supp_multi_parameter_qfim}
\end{equation}
When \(J_{\boldsymbol\theta}(T)\) is nonsingular, the QCRB for estimating \(\theta_1\) is
\begin{equation}
    \operatorname{Var}(\hat\theta_1)
    \geq
    \frac{1}{\nu}
    \left[J_{\boldsymbol\theta}^{-1}(T)\right]_{11}
    =
    \frac{1}{\nu J_{\theta_1|\boldsymbol\theta_{\rm n}}(T)},
\end{equation}
where
\begin{equation}
    J_{\theta_1|\boldsymbol\theta_{\rm n}}(T)
    =
    J_{\theta_1}(T)
    -
    J_{\theta_1\boldsymbol\theta_{\rm n}}(T)
    J_{\boldsymbol\theta_{\rm n}}^{-1}(T)
    J_{\boldsymbol\theta_{\rm n}\theta_1}(T).
    \label{eq:supp_multi_parameter_effective_qfi_schur}
\end{equation}
If the nuisance block is singular, the effective QFI is given by the generalized Schur complement
\begin{equation}
    J_{\theta_1|\boldsymbol\theta_{\rm n}}(T)
    =
    J_{\theta_1}(T)
    -
    J_{\theta_1\boldsymbol\theta_{\rm n}}(T)
    J_{\boldsymbol\theta_{\rm n}}^{+}(T)
    J_{\boldsymbol\theta_{\rm n}\theta_1}(T),
    \label{eq:supp_multi_parameter_effective_qfi_generalized_schur}
\end{equation}
where \({}^{+}\) denotes the Moore--Penrose pseudoinverse.

We now seek a variational representation of this effective QFI in terms of a projected generator. Define the effective generator
\begin{equation}
    S_{\theta_1}^{\rm eff}(\boldsymbol\beta,T)
    =
    S_{\theta_1}(T)
    -
    \sum_{\ell=2}^{N}
    \beta_\ell S_{\theta_\ell}(T),
    \qquad
    \boldsymbol\beta
    =
    (\beta_2,\ldots,\beta_N)^T.
    \label{eq:supp_multi_parameter_effective_generator}
\end{equation}
For a fixed probe \(\ket{\varphi_0}\), we have
\begin{align}
    \Delta^2
    S_{\theta_1}^{\rm eff}(\boldsymbol\beta,T)
    &=
    \Delta^2 S_{\theta_1}(T)
    -
    2\sum_{\ell=2}^{N}
    \beta_\ell
    \operatorname{Cov}
    \left(
        S_{\theta_1}(T),S_{\theta_\ell}(T)
    \right)
   +
    \sum_{\ell,m=2}^{N}
    \beta_\ell\beta_m
    \operatorname{Cov}
    \left(
        S_{\theta_\ell}(T),S_{\theta_m}(T)
    \right).
    \label{eq:supp_multi_parameter_effective_generator_variance}
\end{align}
Using Eq.~\eqref{eq:supp_qfim_generator_form}, this becomes
\begin{equation}
    4\Delta^2
    S_{\theta_1}^{\rm eff}(\boldsymbol\beta,T)
    =
    J_{\theta_1}(T)
    -
    2\boldsymbol\beta^T
    J_{\boldsymbol\theta_{\rm n}\theta_1}(T)
    +
    \boldsymbol\beta^T
    J_{\boldsymbol\theta_{\rm n}}(T)
    \boldsymbol\beta.
    \label{eq:supp_multi_parameter_effective_generator_qfim}
\end{equation}

For a nonsingular nuisance block, the optimal $\beta$ that minimizes this quadratic form is given by
\begin{equation}
    \boldsymbol\beta^\star
    =
    J_{\boldsymbol\theta_{\rm n}}^{-1}(T)
    J_{\boldsymbol\theta_{\rm n}\theta_1}(T),
\label{eq:supp_multi_parameter_beta_probe_dependent}
\end{equation}
and the minimal value recovers the Schur complement
\begin{align}
    J_{\theta_1|\boldsymbol\theta_{\rm n}}(T)
    &=
    4
    \min_{\boldsymbol\beta\in\mathbb R^{N-1}}
    \Delta^2
    S_{\theta_1}^{\rm eff}(\boldsymbol\beta,T)
    =
    J_{\theta_1}(T)
    -
    J_{\theta_1\boldsymbol\theta_{\rm n}}(T)
    J_{\boldsymbol\theta_{\rm n}}^{-1}(T)
    J_{\boldsymbol\theta_{\rm n}\theta_1}(T).
    \label{eq:supp_multi_parameter_variance_schur}
\end{align}
For a singular nuisance block, positivity of the full QFIM  implies the range condition
\begin{equation}
    \left[
        I
        -
        J_{\boldsymbol\theta_{\rm n}}(T)
        J_{\boldsymbol\theta_{\rm n}}^{+}(T)
    \right]
    J_{\boldsymbol\theta_{\rm n}\theta_1}(T)
    =
    0.
    \label{eq:supp_multi_parameter_range_condition}
\end{equation}
A minimizer may therefore be chosen as
\begin{equation}
    \boldsymbol\beta
    =
    J_{\boldsymbol\theta_{\rm n}}^{+}(T)
    J_{\boldsymbol\theta_{\rm n}\theta_1}(T),
    \label{eq:supp_multi_parameter_beta_pseudoinverse}
\end{equation}
up to an arbitrary vector in the null space of \(J_{\boldsymbol\theta_{\rm n}}(T)\). Hence, for both nonsingular and singular nuisance blocks, we have the unified variational representation
\begin{equation}
    J_{\theta_1|\boldsymbol\theta_{\rm n}}(T)
    =
    4
    \min_{\boldsymbol\beta\in\mathbb R^{N-1}}
    \Delta^2
    S_{\theta_1}^{\rm eff}(\boldsymbol\beta,T).
\label{eq:supp_multi_parameter_effective_qfi_projected_generator}
\end{equation}

We now extend the optimal control construction. Define the instantaneous effective velocity
\(    V_{\theta_1}^{\rm eff}(\boldsymbol\beta,t)
    =
    V_{\theta_1}(t)
    -
    \sum_{\ell=2}^{N}
    \beta_\ell V_{\theta_\ell}(t),\)    
and let
\(    \boldsymbol\beta^\star
    =
    \operatorname*{arg\,min}_{\boldsymbol\beta\in\mathbb R^{N-1}}
    \int_0^T
    \left|
        V_{\theta_1}^{\rm eff}(\boldsymbol\beta,t)
    \right|_s
    \mathrm{d}t.\)
Let $\ket{v_{\max}(t)}$ and $\ket{v_{\min}(t)}$ denote the eigenstates
associated with the largest and smallest eigenvalues of
$V_{\theta_1}^{\rm eff}(\boldsymbol{\beta}^{\star},t)$, respectively.
We first consider the nondegenerate case. The optimal initial probe is
\begin{equation}
\ket{\varphi_0^\star}
=
\frac{
\ket{v_{\max}(0)}
+
\ket{v_{\min}(0)}
}{\sqrt{2}} .
\label{eq:supp_multi_parameter_optimal_probe}
\end{equation}
The optimal control is chosen to transport the extremal eigenstates throughout the evolution,
according to
\begin{equation}
U_{\rm tot}(\boldsymbol{\theta}_0,t)
\ket{v_{\max/\min}(0)}
=
\ket{v_{\max/\min}(t)},
\qquad
0\leq t\leq T ,
\label{eq:supp_multi_parameter_transport_control}
\end{equation}
where $\boldsymbol{\theta}_0$ is the operating point.
Under this control, $\ket{v_{\max}(0)}$ and $\ket{v_{\min}(0)}$ remain the
extremal eigenstates of the accumulated effective generator
\begin{equation}
S_{\theta_1}^{\rm eff}(\boldsymbol{\beta}^{\star},T)
=
\int_0^T
U_{\rm tot}^{\dagger}(\boldsymbol{\theta}_0,t)
V_{\theta_1}^{\rm eff}(\boldsymbol{\beta}^{\star},t)
U_{\rm tot}(\boldsymbol{\theta}_0,t)
\mathrm{d}t .
\end{equation}
As a result, the probe $\ket{\varphi_0^\star}$ in
Eq.~\eqref{eq:supp_multi_parameter_optimal_probe} attains the maximal variance of the accumulated effective generator
\begin{align}
4\Delta^2
S_{\theta_1}^{\rm eff}(\boldsymbol{\beta}^{\star},T)
=
\left|
S_{\theta_1}^{\rm eff}(\boldsymbol{\beta}^{\star},T)
\right|_s^2
=
\left[
\int_0^T
\left|
V_{\theta_1}^{\rm eff}(\boldsymbol{\beta}^{\star},t)
\right|_s
\mathrm{d}t
\right]^2,
\label{eq:supp_multi_parameter_width_saturation}
\end{align}
thereby saturating the spectral-width bound.

Next, we show that the scheme is insensitive to the change of nuisance parameters in the first order. Since $\boldsymbol{\beta}^{\star}$ minimizes the integrated spectral width, the
first-order optimality condition, similar to Eq.(\ref{eq:supp_integrated_nuisance_balance}), gives
\begin{equation}
\int_0^T
\left[
\bra{v_{\max}(t)}
V_{\theta_\ell}(t)
\ket{v_{\max}(t)}
-
\bra{v_{\min}(t)}
V_{\theta_\ell}(t)
\ket{v_{\min}(t)}
\right]
\mathrm{d}t
=
0 ,
\qquad
\ell=2,\ldots,N .
\label{eq:supp_multi_parameter_nuisance_balance}
\end{equation}
Using Eq.~\eqref{eq:supp_multi_parameter_transport_control}, this is equivalent to
\begin{equation}
\bra{v_{\max}(0)}
S_{\theta_\ell}(T)
\ket{v_{\max}(0)}
=
\bra{v_{\min}(0)}
S_{\theta_\ell}(T)
\ket{v_{\min}(0)},
\qquad
\ell=2,\ldots,N .
\label{eq:supp_multi_parameter_equal_nuisance_response}
\end{equation}
This equality guarantees that every nuisance generator has identical expectation values on the two extremal states. Combined with the fact that $\ket{v_{\max}(0)}$ and $\ket{v_{\min}(0)}$ remain the eigenstates of $S_{\theta_1}^{\rm eff}(\boldsymbol{\beta}^{\star},T)$, we have 
\begin{equation}
\operatorname{Cov}
\left(
S_{\theta_1}^{\rm eff}(\boldsymbol{\beta}^{\star},T),
S_{\theta_\ell}(T)
\right)
=
0,
\qquad
\ell=2,\ldots,N ,
\label{eq:supp_multi_parameter_covariance_zero}
\end{equation}
for the optimal probe $\ket{\varphi_0^\star}$.
So the effective target generator is orthogonal, in the covariance sense defined by the optimal probe state, to all nuisance generators.

It remains to verify that this $\boldsymbol{\beta}^{\star}$, defined as \(    \boldsymbol\beta^\star
    =
    \operatorname*{arg\,min}_{\boldsymbol\beta\in\mathbb R^{N-1}}
    \int_0^T
    \left|
        V_{\theta_1}^{\rm eff}(\boldsymbol\beta,t)
    \right|_s
    \mathrm{d}t\), is consistent with the minimizer in Eq.(\ref{eq:supp_multi_parameter_effective_qfi_projected_generator}). To show this, let
$\widetilde{\boldsymbol{\beta}}
=
\boldsymbol{\beta}^{\star}
+
\delta\boldsymbol{\beta}$, then
\begin{align}
\Delta^2
S_{\theta_1}^{\rm eff}(\widetilde{\boldsymbol{\beta}},T)
=
\Delta^2
S_{\theta_1}^{\rm eff}(\boldsymbol{\beta}^{\star},T)
+
\sum_{\ell,m=2}^{N}
\delta\beta_\ell\delta\beta_m
\operatorname{Cov}
\left(
S_{\theta_\ell}(T),
S_{\theta_m}(T)
\right)
\geq
\Delta^2
S_{\theta_1}^{\rm eff}(\boldsymbol{\beta}^{\star},T),
\label{eq:supp_multi_parameter_min_attained}
\end{align}
where the linear terms vanish because of
Eq.~\eqref{eq:supp_multi_parameter_covariance_zero}, and the last inequality
follows from the positive semidefiniteness of the nuisance covariance
matrix. It is therefore consistent.

Combining them together, we then have
\begin{equation}
J_{\theta_1|\boldsymbol{\theta}_{\rm n}}^{\star}(T)
=
\left|
S_{\theta_1}^{\rm eff}(\boldsymbol{\beta}^{\star},T)
\right|_s^2
=
\left[
\int_0^T
\left|
V_{\theta_1}^{\rm eff}(\boldsymbol{\beta}^{\star},t)
\right|_s
\mathrm{d}t
\right]^2 .
\label{eq:supp_multi_parameter_optimal_qfi}
\end{equation}
Hence the probe and control constructed above attain the optimal effective QFI in the presence of an arbitrary number of nuisance parameters.

We now construct the measurement that achieves this bound. Define 
\begin{equation}
O_{\theta_1}^{\rm eff}
=
-i
\ket{v_{\max}(0)}
\bra{v_{\min}(0)}
+
i
\ket{v_{\min}(0)}
\bra{v_{\max}(0)} .
\label{eq:supp_multi_parameter_optimal_observable}
\end{equation}
For the optimal probe, it satisfies the condition required for optimal observable \cite{hou2021super}
\begin{equation}
\left[
S_{\theta_1}^{\rm eff}(\boldsymbol{\beta}^{\star},T)
-
\left\langle
S_{\theta_1}^{\rm eff}(\boldsymbol{\beta}^{\star},T)
\right\rangle
\right]
\ket{\varphi_0^\star}
=
i
\frac{
\left|
S_{\theta_1}^{\rm eff}(\boldsymbol{\beta}^{\star},T)
\right|_s
}{2}
O_{\theta_1}^{\rm eff}
\ket{\varphi_0^\star},
\label{eq:supp_multi_parameter_optimal_observable_relation}
\end{equation}
Hence, the eigenbasis of $O_{\theta_1}^{\rm eff}$ forms an optimal measurement. In the Schr\"odinger picture, the corresponding measurement
basis is
\begin{equation}
\ket{\pi_\pm}
=
U_{\rm tot}(\boldsymbol{\theta}_0,T)
\frac{
\ket{v_{\max}(0)}
\pm
i\ket{v_{\min}(0)}
}{\sqrt{2}}
=
\frac{
\ket{v_{\max}(T)}
\pm
i\ket{v_{\min}(T)}
}{\sqrt{2}} .
\label{eq:supp_multi_parameter_optimal_measurement}
\end{equation}
A binary POVM can be constructed as
\begin{equation}
M_\pm
=
\ket{\pi_\pm}\bra{\pi_\pm}
+
\frac{1}{2}
\left(
\mathbb I
-
\ket{\pi_+}\bra{\pi_+}
-
\ket{\pi_-}\bra{\pi_-}
\right),
\label{eq:supp_multi_parameter_povm}
\end{equation}
which satisfies $M_++M_-=\mathbb I$. The probabilities of the measurement outcome are given by
\begin{equation}
p_\pm(\boldsymbol{\theta})
=
\bra{\varphi_0^\star}
U_{\rm tot}^{\dagger}(\boldsymbol{\theta},T)
M_\pm
U_{\rm tot}(\boldsymbol{\theta},T)
\ket{\varphi_0^\star},
\label{eq:supp_multi_parameter_probabilities}
\end{equation}
with
$p_\pm(\boldsymbol{\theta}_0)=1/2$. Eq.~\eqref{eq:supp_multi_parameter_equal_nuisance_response}, immediately implies that the measurement is locally insensitive to every nuisance parameter:
\begin{equation}
\left.
\partial_{\theta_\ell}
(p_+-p_-)
\right|_{\boldsymbol{\theta}_0}
=
0,
\qquad
\ell=2,\ldots,N .
\label{eq:supp_multi_parameter_nuisance_response_zero}
\end{equation}
For the target parameter, using the decomposition
\begin{equation}
S_{\theta_1}(T)
=
S_{\theta_1}^{\rm eff}(\boldsymbol{\beta}^{\star},T)
+
\sum_{\ell=2}^{N}
\beta_\ell^\star S_{\theta_\ell}(T),
\label{eq:supp_multi_parameter_target_decomposition}
\end{equation}
so that the nuisance contributions to the target response cancel according
to Eq.~\eqref{eq:supp_multi_parameter_equal_nuisance_response}.
Following the same calculation as in the single-nuisance case, the derivative of the probability difference with respect to any parameter can be expressed in terms of the generators as
\begin{equation}
\left.
\partial_{\theta_j}(p_+-p_-)
\right|_{\boldsymbol{\theta}_0}
=
\frac{4}{
\left|
S_{\theta_1}^{\rm eff}(\boldsymbol{\beta}^{\star},T)
\right|_s
}
\operatorname{Cov}\left(
S_{\theta_1}^{\rm eff}(\boldsymbol{\beta}^{\star},T),
S_{\theta_j}(T)
\right),
\qquad
j=1,\ldots,N .
\label{eq:supp_multi_parameter_probability_derivative}
\end{equation}
For the target parameter, using Eq.~\eqref{eq:supp_multi_parameter_target_decomposition} together with the vanishing nuisance covariances, we have
\begin{align}
\operatorname{Cov}\left(
S_{\theta_1}^{\rm eff},
S_{\theta_1}
\right)
&=
\operatorname{Cov}\left[
S_{\theta_1}^{\rm eff},
(S_{\theta_1}^{\rm eff}
+
\sum_{\ell=2}^{N}
\beta_\ell^\star S_{\theta_\ell})
\right]
=
\Delta^2 S_{\theta_1}^{\rm eff},
\end{align}
where the arguments are omitted for compactness. Since the optimal probe satisfies
\(
4\Delta^2 S_{\theta_1}^{\rm eff}
=
|S_{\theta_1}^{\rm eff}|_s^2
\),
Eq.~\eqref{eq:supp_multi_parameter_probability_derivative} gives
\begin{equation}
\left.
\partial_{\theta_1}
(p_+-p_-)
\right|_{\boldsymbol{\theta}_0}
=
\left|
S_{\theta_1}^{\rm eff}(\boldsymbol{\beta}^{\star},T)
\right|_s .
\label{eq:supp_multi_parameter_target_response}
\end{equation}
Together with Eq.~\eqref{eq:supp_multi_parameter_nuisance_response_zero}, this yields
\begin{equation}
p_\pm(
\boldsymbol{\theta}_0+\delta\boldsymbol{\theta}
)
=
\frac{1}{2}
\pm
\frac{1}{2}
\left|
S_{\theta_1}^{\rm eff}(\boldsymbol{\beta}^{\star},T)
\right|_s
\delta\theta_1
+
O\!\left(
\|\delta\boldsymbol{\theta}\|^2
\right).
\label{eq:supp_multi_parameter_probability_expansion}
\end{equation}
Thus, to first order, the measurement response depends exclusively on the target parameter while retaining the full effective sensitivity.

Since \(p_\pm(\boldsymbol{\theta}_0)=1/2\), the resulting classical Fisher information matrix is
\begin{equation}
F(\boldsymbol{\theta}_0,T)
=
\begin{pmatrix}
\left|
S_{\theta_1}^{\rm eff}(\boldsymbol{\beta}^{\star},T)
\right|_s^2
&
\boldsymbol{0}^{T}
\\
\boldsymbol{0}
&
\boldsymbol{0}
\end{pmatrix}.
\label{eq:supp_multi_parameter_cfim}
\end{equation}
The generalized Schur complement based on the Moore--Penrose pseudoinverse therefore gives
\begin{equation}
F_{\theta_1|\boldsymbol{\theta}_{\rm n}}(T)
=
\left|
S_{\theta_1}^{\rm eff}(\boldsymbol{\beta}^{\star},T)
\right|_s^2
=
J_{\theta_1|\boldsymbol{\theta}_{\rm n}}^{\star}(T).
\label{eq:supp_multi_parameter_cfim_qfim_equality}
\end{equation}
Hence, the binary measurement achieves the optimal effective QFI for the target parameter while having vanishing first-order response to all nuisance parameters.

To convert the binary measurement outcomes into an estimator for the target parameter, we introduce the random variable
\begin{equation}
    Y
    =
    \begin{cases}
        +1, & \text{if }M_+\text{ occurs},\\
        -1, & \text{if }M_-\text{ occurs}.
    \end{cases}
    \label{eq:Y_definition}
\end{equation}
Using Eq.~\eqref{eq:supp_multi_parameter_probability_expansion}, its expectation value near the operating point is
\begin{equation}
    \mathbb E_{\boldsymbol\theta}[Y]
    =
    p_+-p_-
    =
    \left|
        S_{\theta_1}^{\rm eff}(\boldsymbol\beta^\star,T)
    \right|_s
    (\theta_1-\theta_{0,1})
    +
    O\!\left(\|\delta\boldsymbol\theta\|^2\right),
    \label{eq:Y_expectation}
\end{equation}
where \(\boldsymbol\theta_0=(\theta_{0,1},\ldots,\theta_{0,N})^T\). Thus, to first order, the expectation value of \(Y\) is sensitive only to the target parameter and varies linearly with its deviation from the operating point.

For \(\nu\) independent repetitions, let
\begin{equation}
    \overline Y
    =
    \frac{1}{\nu}
    \sum_{r=1}^{\nu}Y_r
    \label{eq:Y_sample_mean}
\end{equation}
denote the sample mean. An explicit locally unbiased estimator can be obtained as 
\begin{equation}
    \hat\theta_1
    =
    \theta_{0,1}
    +
    \frac{
        \overline Y
    }{
        \left|
            S_{\theta_1}^{\rm eff}(\boldsymbol\beta^\star,T)
        \right|_s
    }.
    \label{eq:target_estimator}
\end{equation}
Equivalently, if \(n_+\) and \(n_-\) are the observed outcome counts, with \(n_++n_-=\nu\), then
\begin{equation}
    \hat\theta_1
    =
    \theta_{0,1}
    +
    \frac{
        n_+-n_-
    }{
        \nu
        \left|
            S_{\theta_1}^{\rm eff}(\boldsymbol\beta^\star,T)
        \right|_s
    }.
    \label{eq:count_estimator}
\end{equation}
At the operating point,
\begin{equation}
    \mathbb E_{\boldsymbol\theta_0}[\hat\theta_1]
    =
    \theta_{0,1},
\end{equation}
and
\begin{align}
    \left.
    \partial_{\theta_1}
    \mathbb E_{\boldsymbol\theta}[\hat\theta_1]
    \right|_{\boldsymbol\theta_0}
    &=
    1,
    \nonumber\\
    \left.
    \partial_{\theta_\ell}
    \mathbb E_{\boldsymbol\theta}[\hat\theta_1]
    \right|_{\boldsymbol\theta_0}
    &=
    0,
    \qquad
    \ell=2,\ldots,N.
    \label{eq:supp_multi_parameter_local_unbiasedness}
\end{align}
Hence, \(\hat\theta_1\) is locally unbiased for the target parameter and locally insensitive to all nuisance directions. Since \(Y^2=1\), we have
\begin{equation}
    \operatorname{Var}_{\boldsymbol\theta_0}(\hat\theta_1)
    =
    \frac{1}{
        \nu
        \left|
            S_{\theta_1}^{\rm eff}(\boldsymbol\beta^\star,T)
        \right|_s^2
    }
    =
    \frac{1}{
        \nu
        J_{\theta_1|\boldsymbol\theta_{\rm n}}^{\star}(T)
    }.
    \label{eq:estimator_variance_optimal}
\end{equation}
The estimator therefore saturates the local precision bound.

\section{Robustness of the adaptive protocol}

We now illustrate the robustness of the adaptive protocol using the examples presented in the manuscript. 

For the linear-drift model presented in the End Matter,
\begin{equation}
    H(t)
    =
    \frac{B}{2}
    (\omega t+\alpha)\sigma_z,
\end{equation}
where \(\omega\) is the target parameter and \(\alpha\) is a nuisance
parameter. The parameter velocities are
\begin{equation}
    V_\omega(t)
    =
    \partial_\omega H(t)=
    \frac{B}{2}t\,\sigma_z,
    \qquad
    V_\alpha(t)=
    \partial_\alpha H(t)
    =
    \frac{B}{2}\sigma_z.
\end{equation}
For a real coefficient \(\beta\), the effective parameter velocity is
\begin{equation}
    V_\omega^{\rm eff}(\beta,t)
    =
    V_\omega(t)-\beta V_\alpha(t)
    =
    \frac{B}{2}(t-\beta)\sigma_z .
    \label{eq:supp_linear_effective_velocity}
\end{equation}
The optimal coefficient is therefore
\begin{equation}
    \beta^\star
    =
    \arg\min_\beta
    |B|
    \int_0^T |t-\beta|\,\mathrm{d}t
    =
    \frac{T}{2}.
    \label{eq:supp_linear_beta}
\end{equation}

Let \(U_c(t)\) denote the control propagator generated by a sequence of
instantaneous \(\pi\)-pulses, and define the toggling function
\(y(t)=\pm1\) by
\begin{equation}
    U_c^\dagger(t)\sigma_z U_c(t)
    =
    y(t)\sigma_z .
\end{equation}
The controlled generators are
\begin{equation}
    S_i(T)
    =
    \int_0^T
    y(t)V_i(t)\,\mathrm{d}t,
    \qquad
    i=\omega,\alpha,
\end{equation}
and the effective generator is
\begin{equation}
    S_\omega^{\rm eff}(\beta,T)
    =
    S_\omega(T)-\beta S_\alpha(T)
    =
    \frac{B}{2}
    \int_0^T
    y(t)(t-\beta)\,\mathrm{d}t\,\sigma_z.
    \label{eq:supp_linear_effective_generator}
\end{equation}
The optimal control is
\begin{equation}
    y^\star(t)
    =
    \operatorname{sgn}
    \left(t-\frac{T}{2}\right),
    \label{eq:supp_linear_optimal_control}
\end{equation}
which can be implemented by a single instantaneous \(\pi\) pulse at
\(t=T/2\). With this control,
\begin{equation}
    S_\alpha(T)
    =
    \frac{B}{2}
    \int_0^T y^\star(t)\,\mathrm{d}t\,\sigma_z
    =
    0,
\end{equation}
while
\begin{align}
    S_\omega^{\rm eff}(\beta^\star,T)
    &=
    \frac{B}{2}
    \int_0^T
    \left|t-\frac{T}{2}\right|
    \mathrm{d}t\,\sigma_z=
    \frac{BT^2}{8}\sigma_z.
    \label{eq:supp_linear_optimal_generator}
\end{align}
Its spectral width is therefore
\begin{equation}
    \left|
    S_\omega^{\rm eff}(\beta^\star,T)
    \right|_s
    =
    \frac{|B|T^2}{4},
\end{equation}
giving
\begin{equation}
    J_{\omega|\alpha}^{\star}(T)
    =
    \frac{B^2T^4}{16}.
    \label{eq:supp_linear_optimal_information}
\end{equation}

In an adaptive implementation, replacing the unknown parameters by
estimates \((\hat{\omega},\hat{\alpha})\) leaves both the optimal
coefficient and the control unchanged:
\begin{equation}
    \hat{\beta}^\star
    =
    \frac{T}{2},
    \qquad
    \hat{y}(t)
    =
    \operatorname{sgn}
    \left(t-\frac{T}{2}\right).
\end{equation}
Hence the implemented effective generator remains
\begin{equation}
    \widetilde{S}_\omega^{\rm eff}(T)
    =
    \frac{BT^2}{8}\sigma_z
\end{equation}
for arbitrary estimation errors in \(\omega\) and \(\alpha\), and
\begin{equation}
    \frac{
        J_{\omega|\alpha}^{\rm ad}(T)
    }{
        J_{\omega|\alpha}^{\star}(T)
    }
    =
    1.
    \label{eq:supp_linear_robustness}
\end{equation}
Thus, the adaptive protocol attains the optimal nuisance-aware
information exactly for this model, since the optimal control is
independent of the unknown parameter values.

Now we consider the
frequency-estimation model presented in the End Matter,
\begin{equation}
    H(t)
    =
    \frac{B}{2}
    \cos(\omega t+\phi)\sigma_z,
\end{equation}
where \(\omega\) is the target parameter and \(\phi\) is a nuisance
phase. The parameter velocities are
\begin{align}
    V_\omega(t)
    &=
    \partial_\omega H(t)
    =
    -\frac{B}{2}
    t\sin(\omega t+\phi)\sigma_z,
    \\
    V_\phi(t)
    &=
    \partial_\phi H(t)
    =
    -\frac{B}{2}
    \sin(\omega t+\phi)\sigma_z.
\end{align}
For a real coefficient \(\beta\), the effective parameter velocity for
estimating \(\omega\) in the presence of the nuisance phase is
\begin{align}
    V_\omega^{\rm eff}(\beta,t)
    &=
    V_\omega(t)-\beta V_\phi(t)
    =
    -\frac{B}{2}
    (t-\beta)
    \sin(\omega t+\phi)\sigma_z .
    \label{eq:supp_adaptive_effective_velocity}
\end{align}
The optimal coefficient is
\begin{equation}
    \beta^\star
    =
    \arg\min_{\beta}
    \int_0^T
    \left|
        V_\omega^{\rm eff}(\beta,t)
    \right|_s
    \mathrm{d}t
    =
    \arg\min_{\beta}
    B
    \int_0^T
    |t-\beta|
    |\sin(\omega t+\phi)|
    \mathrm{d}t ,
    \label{eq:supp_adaptive_beta_opt}
\end{equation}
which is the weighted median of \(t\) with weight
\(|\sin(\omega t+\phi)|\).

The controlled generators associated with the two parameters are
\begin{equation}
    S_i(T)
    =
    \int_0^T
    U_c^\dagger(t)
    V_i(t)
    U_c(t)
    \mathrm{d}t,
    \qquad
    i=\omega,\phi.
\end{equation}
Accordingly, the effective generator is
\begin{align}
    S_\omega^{\rm eff}(\beta,T)
    &=
    S_\omega(T)-\beta S_\phi(T)
    =
    \int_0^T
    U_c^\dagger(t)
    V_\omega^{\rm eff}(\beta,t)
    U_c(t)
    \mathrm{d}t .
    \label{eq:supp_adaptive_effective_generator}
\end{align}

The optimal control is
\begin{equation}
    y^\star(t)
    =
    \operatorname{sgn}
    \left[
        (t-\beta^\star)
        \sin(\omega t+\phi)
    \right].
    \label{eq:supp_adaptive_optimal_control}
\end{equation}
With this choice, all contributions to the effective generator are
aligned, giving
\begin{equation}
    S_\omega^{\rm eff}(\beta^\star,T)
    =
    -\frac{B}{2}
    \int_0^T
    |t-\beta^\star|
    |\sin(\omega t+\phi)|
    \mathrm{d}t\,
    \sigma_z .
\end{equation}
The resulting optimal effective information is therefore
\begin{equation}
    J_{\omega|\phi}^{\star}(T)
    =
    \left[
        B
        \int_0^T
        |t-\beta^\star|
        |\sin(\omega t+\phi)|
        \mathrm{d}t
    \right]^2 .
    \label{eq:supp_adaptive_optimal_information}
\end{equation}
In the long-time regime,
\(\beta^\star\simeq T/2\), and averaging over the rapid oscillations
gives
\begin{equation}
    J_{\omega|\phi}^{\star}(T)
    \simeq
    \frac{B^2T^4}{4\pi^2}.
    \label{eq:supp_adaptive_long_time_optimum}
\end{equation}

In an adaptive implementation, the unknown parameters in the control
are replaced by their current estimates
\((\hat{\omega},\hat{\phi})\). Thus,
\begin{equation}
    \hat{\beta}^{\star}
    =
    \arg\min_{\beta}
    B
    \int_0^T
    |t-\beta|
    |\sin(\hat{\omega}t+\hat{\phi})|
    \mathrm{d}t ,
\end{equation}
and the implemented toggling function is
\begin{equation}
    \hat{y}(t)
    =
    \operatorname{sgn}
    \left[
        (t-\hat{\beta}^{\star})
        \sin(\hat{\omega}t+\hat{\phi})
    \right].
    \label{eq:supp_adaptive_control}
\end{equation}
We denote the estimation errors by
\begin{equation}
    \delta\omega
    =
    \hat{\omega}-\omega,
    \qquad
    \delta\phi
    =
    \hat{\phi}-\phi.
\end{equation}

For the regime \(|\omega|T\gg1\), \(|\delta\phi|\ll1\), and \(|\delta\omega|T\ll1\), both \(\beta^\star\) and \(\hat{\beta}^\star\) approach \(T/2\), while the accumulated control mismatch remains small over the interrogation time. Averaging over the rapid oscillations then gives
\begin{equation}
    \hat{y}(t)\sin(\omega t+\phi)
    \simeq
    \frac{2}{\pi}
    \operatorname{sgn}
    \left(t-\frac{T}{2}\right)
    \cos(\delta\phi+\delta\omega t).
    \label{eq:supp_adaptive_carrier_average}
\end{equation}
The effective generator obtained with the mismatched control is
therefore
\begin{align}
    \widetilde{S}_\omega^{\rm eff}(T)
    &=
    \int_0^T
    \hat{y}(t)
    V_\omega^{\rm eff}(\hat{\beta}^\star,t)
    \mathrm{d}t
    \nonumber\\
    &\simeq
    -\frac{B}{\pi}
    \int_0^T
    \left|t-\frac{T}{2}\right|
    \cos(\delta\phi+\delta\omega t)
    \mathrm{d}t\,
    \sigma_z
    \nonumber\\
    &\simeq
    -\frac{BT^2}{4\pi}
    \left[
        1
        -
        \frac{(\delta\phi)^2}{2}
        -
        \frac{\delta\phi\,\delta\omega T}{2}
        -
        \frac{3(\delta\omega T)^2}{16}
    \right]\sigma_z .
    \label{eq:supp_adaptive_mismatched_generator}
\end{align}
The corresponding effective information relative to the local optimum is
\begin{align}
    \frac{
        J_{\omega|\phi}^{\rm ad}(T)
    }{
        J_{\omega|\phi}^{\star}(T)
    }
    &\simeq
    1
    -
    (\delta\phi)^2
    -
    \delta\phi\,\delta\omega T
    -
    \frac{3}{8}(\delta\omega T)^2
    =
    1
    -
    \left(
        \delta\phi
        +
        \frac{\delta\omega T}{2}
    \right)^2
    -
    \frac{(\delta\omega T)^2}{8}.
    \label{eq:supp_adaptive_mismatch_result}
\end{align}
Thus the loss caused by imperfect estimates of \(\omega\) and \(\phi\)
is second order in \(\delta\phi\) and \(\delta\omega T\), and the
adaptive interrogation is locally insensitive to estimation errors to
first order.

\end{document}